\documentclass{jfm}
\usepackage{graphicx}
\usepackage{amsmath}
\usepackage{color}
\usepackage{xfrac}
\usepackage{algorithm}
\usepackage[noend]{algpseudocode}
\usepackage{gensymb}

\usepackage{pifont}
\usepackage{amssymb}
\usepackage{wasysym}

\shorttitle{Complex sliding flows of yield-stress fluids}
\shortauthor{Chaparian and Tammisola}

\title{Complex sliding flows of yield-stress fluids}

\author{Emad Chaparian\corresp{\email{emadc@mech.kth.se}}
 \and Outi Tammisola}

\affiliation{Linn{\'e} FLOW Centre and SeRC, Department of Engineering Mechanics, Royal Institute of Technology (KTH), SE-100 44, Stockholm, Sweden}

\begin{document}

\maketitle

\begin{abstract}
A theoretical and numerical study of complex sliding flows of yield-stress fluids is presented. Yield-stress fluids are known to slide over solid surfaces if the tangential stress exceeds the {\it sliding yield stress}. The sliding may occur due to various microscopic phenomena such as the formation of a infinitesimal lubrication layer of the solvent and/or elastic deformation of the suspended soft particles in the vicinity of the solid surfaces. This leads to a `stick-slip' law which complicates the modelling and analysis of the hydrodynamic characteristics of the yield-stress fluid flow. In the present study, we formulate the problem of sliding flow beyond one-dimensional rheometric flows. Then, a numerical scheme based on the augmented Lagrangian method is presented to attack these kind of problems. Theoretical tools are developed for analysing the flow/no-flow limit. The whole framework is benchmarked in planar Poiseuille flow and validated against analytical solutions. Then two more complex physical problems are investigated: slippery particle sedimentation and pressure-driven sliding flow in porous media. The {\it yield limit} is addressed in detail for both flow cases. In the particle sedimentation problem, method of characteristics---slipline method---in the presence of slip is revisited from the perfectly-plastic mechanics and used as a helpful tool in addressing the yield limit. Finally, flows through model and randomized porous media are studied. The randomized configuration is chosen to capture more sophisticated aspects of the yield-stress fluid flows in porous media at the yield limit---channelization.
\end{abstract}


\section{Introduction}

Slip behaviour in complex fluids has been attributed to various microscopic phenomena such as chain detachment/desorption at the polymer-wall interface in polymer melts \citep{hatzikiriakos2012wall}, migration/depletion of disperse phase away from the vicinity of the solid boundaries in suspensions \citep{vand1948viscosity}, and elastic deformation of the suspended soft particles lying on the smooth solid boundaries \citep{meeker2004PRL} in pastes---yield-stress fluids. In the present study, however, we rather investigate the macroscopic consequences of slip on hydrodynamic features of yield-stress fluid flows, in order to form a bridge over these two distinct scales.

In experiments, roughened (e.g. using sandpapers) or chemically treated surfaces are usually used to suppress slip in measurements of yield-stress fluid flows. For instance, \cite{christel2012stick} proposed a Polymethyl methacrylate (PMMA) treatment to reduce slip by exciting positive surface charges providing electrostatic interactions and squashing the lubricating layer. This is important since in a great number of laboratory experiments, PMMA is used due to its transparency properties which enable flow visualizations. In industrial scales/processes, however, chemical treatments are not always feasible, and therefore it is crucial to analyse the hydrodynamic consequences of sliding flows. Moreover, slip alters rheometric measurements, causing rheological properties of the samples to be inaccurately evaluated, sometimes even by one order of magnitude \citep{poumaere2014unsteady}. This issue has been discussed extensively in the literature, for instance recently by \cite{medina2017tangential,medina2019rheo}. Hence, it would be constructive to have a generic analysis framework of sliding flows in yield-stress fluids, and this is the aim of the present paper. 

In the present study, we formulate the well-known `stick-slip' law for yield-stress fluids beyond 1D rheometric flows and then generalise the previously proposed numerical algorithm \citep{roquet2008adaptive,muravleva2018squeeze} based on the augmented Lagrangian method for attacking this kind of problems. Theoretical tools are developed in the presence of slip. The whole framework is first benchmarked by a simple channel Poiseuille flow. Then we proceed to investigate flows in complex geometries: the creeping flows about a circular cylinder and the pressure-driven flows in model and randomized porous media.

\subsection{General slip law for yield-stress fluids}
Characterising the slip behaviour of yield-stress fluids has been the main objective of a large number of studies from theoretical attempts to experimental measurements. \cite{meeker2004PRL} and \cite{piau2007carbopol} tried to theoretically describe the slip origin in soft particle pastes/microgels from an elastohydrodynamic perspective. Rheometric tests with slippery plates in the cone-plate/plate-plate geometries have been conducted to quantify the slip behaviours of yield-stress fluids ranging from Carbopol gels to emulsions \citep{meeker2004JOR,meeker2004PRL,poumaere2014unsteady,zhang2018wall}. Moreover, very recently, in a series of experiments using Optical coherence tomography (OCT) and particle tracking velocimetry (PTV), \cite{daneshi2019characterising} characterised the slip behaviour of Carbopol gel (a model `simple' yield-stress fluid) inside a capillary tube. The sliding characteristics of yield-stress fluids observed in all the mentioned 1D experimental studies can be summarized in a general slip law as,
\begin{equation}\label{1DSlidingLaw}
\hat{u}_{{s}} = \left\{
\begin{array}{ll}
\hat{\beta}_s \left( \hat{\tau}_{w} - \hat{\tau}_{s} \right)^k, & \text{iff}~~ \vert \hat{\tau}_{w} \vert > \hat{\tau}_s, \\[2pt]
0, & \text{iff}~~ \vert \hat{\tau}_{w} \vert \leqslant \hat{\tau}_s,
\end{array} \right.
\end{equation}
usually termed as `stick-slip' law, where $\hat{u}_{{s}}$ is the slip velocity on the solid surface and $\hat{\tau}_w$ is the shear stress at the solid boundaries. The sliding threshold is called the {\it sliding} yield stress---$\hat{\tau}_s$. The slip coefficient and the power index are designated by $\hat{\beta}_s$ and $k$, respectively. Hence, $\hat{\beta}_s$ has the dimension of $m^{2k+1}/N^k.s$ or $m/Pa^k.s$. Its physical interpretation (for the case $k=1$) is the {\it slip length} over the local {\it effective viscosity} of the fluid, $\hat{\ell}_s / \hat{\mu}_{eff}$. In the entire paper, quantities with a `hat' symbol ($\hat{\cdot}$) are dimensional and others are dimensionless.

\begin{figure}
\centerline{\includegraphics[width=\linewidth]{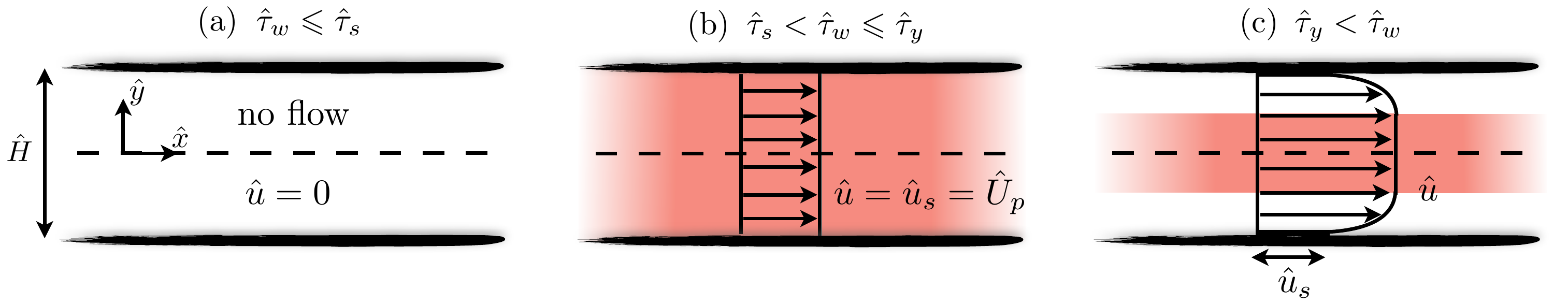}}
\caption{Schematic of the sliding channel flow of a yield-stress fluid; the $x$-axis is aligned and put at the axial centreline of the channel which is formed by the two infinite parallel plates separated by the gap $\hat{H}$ and $y$-axis is in the wall-normal direction: (a) when $\hat{\tau}_w \leqslant \hat{\tau}_s$ there is no flow, ~(b) when $\hat{\tau}_s < \hat{\tau}_w \leqslant \hat{\tau}_y$ then the fluid in the whole gap slides as an unyielded plug, ~(c) when $\hat{\tau}_y < \hat{\tau}_w$ then the fluid in the vicinity of the walls ($\hat{y}_p <\hat{y}$ where $\hat{\tau}_y < \hat{\tau}_{xy} $) yields and at the same time slides over the walls. There is a core unyielded region as well ($\hat{y} \leqslant \hat{y}_p$).}
\label{fig:Schematic_slip}
\end{figure}

For a deeper understanding of the physical meaning of the slip law (\ref{1DSlidingLaw}), we consider the simple unidirectional channel Poiseuille flow in the presence of slip of a viscoplastic fluid,
\begin{equation}\label{const}
  \left\{
    \begin{array}{ll}
      \hat{\tau}_{xy} = \hat{K} \displaystyle \frac{\text{d} \hat{u}}{\text{d} \hat{y}} \left[ \text{abs}\left( \frac{\text{d} \hat{u}}{\text{d} \hat{y}} \right) \right]^{n-1} + \hat{\tau}_y ~\text{sgn}\left( \frac{\text{d} \hat{u}}{\text{d} \hat{y}} \right) & \mbox{iff}\quad \vert \hat{\tau}_{xy} \vert > \hat{\tau}_y, \\[2pt]
      \displaystyle \frac{\text{d} \hat{u}}{\text{d} \hat{y}} = 0 & \mbox{iff}\quad \vert \hat{\tau}_{xy} \vert \leqslant \hat{\tau}_y,
  \end{array} \right.
\end{equation}
where $\hat{K} (Pa \cdot s^n)$ is the consistency of the fluid and $n$ the power-index (Herschel-Bulkley model), $\hat{u}$ the streamwise velocity, $\hat{\tau}_y$ the material's yield stress, $\hat{\tau}_{xy}$ the shear stress, and $\text{sgn}(\cdot)$ and $\text{abs}(\cdot)$ are the {\it sign} and {\it absolute value} functions. Figure \ref{fig:Schematic_slip} represents the slip law schematically. When the applied pressure gradient is small enough, the wall shear stress ($\hat{\tau}_w$) lies below the sliding yield stress (panel (a)) and the slip velocity is zero. Since $\hat{\tau}_s$ is always less than the material's yield stress, then there is no flow in the channel. When the pressure gradient is increased beyond $(\Delta \hat{p}/\hat{L})_c$ where subscript `$_c$' indicates the critical value, the wall shear stress grows beyond $\hat{\tau}_s$ and the fluid slides over the walls. If the wall shear stress is still less than $\hat{\tau}_y$, then the material moves as a sliding unyielded plug with a constant velocity in the entire gap (see panel (b)). When the pressure gradient is increased, at some point, the wall shear stress exceeds the fluid yield stress and the sheared/yielded regions appear, which at the same time slide over the walls; see panel (c). Yet, in the vicinity of the centreline, the shear stress drops below the yield stress and a core unyielded region is formed. The flows corresponding to panels (b) and (c) are usually termed the fully plugged regime and the deformation regime, respectively.

In the present study, the objective is to systematically address the effect of slip on yield-stress fluid flows and its consequences on the yield limit; not only by conducting numerical simulations, but also by forming a theoretical framework for analysing this type of problems. Although some attempts have been made to solve simple rheometric flows in the previous years \citep{philippou2016viscoplastic,panaseti2017viscoplastic,damianou2019viscoplastic}, still the lack of generic numerical implementation frameworks (especially non-regularised methods) and the absence of adequate theoretical frameworks to attack sliding yield-stress fluid flows are discernible. Hence, in what follows, we aim to partially fill this gap in the literature. Beyond that, we study complex sliding flows such as particle sedimentation and pressure-driven flows in porous media within these frameworks.

The outline of the paper is as follows. In \S 2, we set out the slip law beyond 1D for the yield-stress fluid flows and generalise the numerical algorithm previously proposed for simple 1D problems. We then benchmark the numerical algorithm for simple channel Poiseuille flow in \S 3 and derive some theoretical tools for investigating the sliding flows. This will be followed by addressing the sliding flow about a circular particle and the aspects of the particle yield limit in \S4. The next two sections are devoted to the sliding flows in porous media: in \S5 some general hints are made by studying the flow through model porous media and in \S6 some deeper investigations are carried out by performing numerical simulations in randomized porous media. Finally, some conclusions are drawn in the closing section \S7.

\section{Slip law and the numerical algorithm beyond 1D}\label{sec:ps}

In this section, we set out and formulate a sample problem of sliding flow of a yield-stress fluid and generalise the slip law and numerical algorithm for solving such problems beyond 1D. In what follows, we mainly assume that $n=1$ (Bingham model). Later in Appendix \ref{sec:kneq1}, we will address how the present framework can be expanded to encompass the other `simple' yield-stress rheological models, e.g.~Herschel-Bulkley. However, the majority of the physical discussions (e.g.~the yield limit) in the next sections are independent of the value of $n$. We will come back to this point in \S \ref{sec:SD}.

\begin{figure}
\centerline{\includegraphics[width=0.35\linewidth]{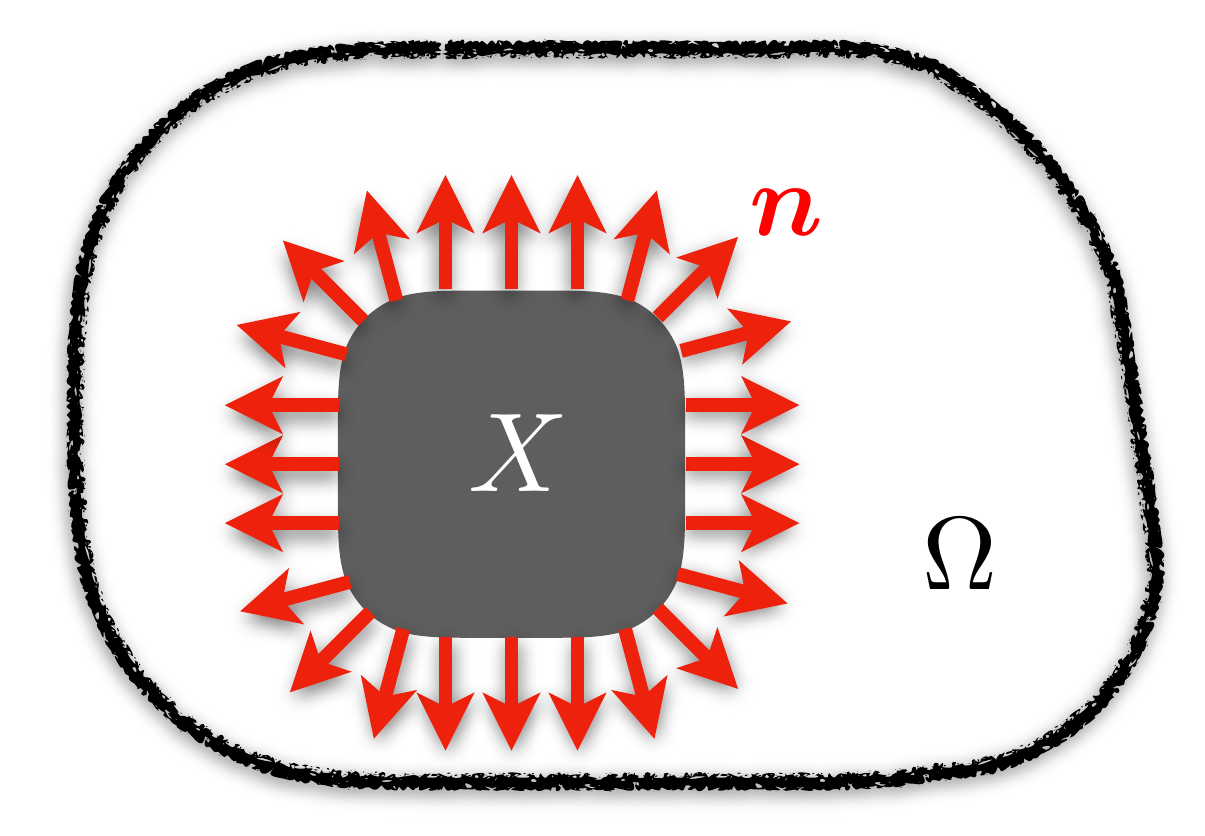}}
\caption{Schematic of the problem $\mathbb{P}$. The full domain is denoted by $\Omega$, and the solid object on which the flow slides in designated by $X$. The red arrows show the outward unit normal vectors of the solid surface $\partial X$. The boundary of the full domain is $\partial \Omega$.}
\label{fig:Schematic_alg}
\end{figure}

{\bf Problem $\mathbb{P}$}: we consider the Stokes flow of a Bingham fluid in $\Omega \setminus \bar{X}$ (see figure \ref{fig:Schematic_alg}):

\begin{equation}
- \boldsymbol{\nabla} \hat{p} + \boldsymbol{\nabla} \boldsymbol{\cdot} \hat{\boldsymbol{\tau}} + \hat{\rho} \hat{\boldsymbol{f}} = 0,
\end{equation}
\begin{equation}\label{const}
  \left\{
    \begin{array}{ll}
      \hat{\boldsymbol{\tau}} = \left( \hat{\mu} + \displaystyle{\frac{\hat{\tau}_y}{\Vert \hat{\dot{\boldsymbol{\gamma}}} \Vert}} \right) \hat{\dot{\boldsymbol{\gamma}}} & \mbox{iff}\quad \Vert \hat{\boldsymbol{\tau}} \Vert > \hat{\tau}_y, \\[2pt]
      \hat{\dot{\boldsymbol{\gamma}}} = 0 & \mbox{iff}\quad \Vert \hat{\boldsymbol{\tau}} \Vert \leqslant \hat{\tau}_y,
  \end{array} \right.
\end{equation}
where $\hat{p}$ is the pressure, $\hat{\boldsymbol{\tau}}$ the deviatoric stress tensor, $\hat{\rho} \hat{\boldsymbol{f}}$ the body force, $\hat{\rho}$ the fluid density, and $\hat{\dot{\boldsymbol{\gamma}}}$ is the rate of deformation tensor. Without loss of generality, we consider no-slip \& no-penetration boundary condition on $\partial \Omega$ as $\hat{\boldsymbol{u}} = \hat{\boldsymbol{u}}_0$ and the slip boundary condition,
\begin{equation}\label{SlidingLaw}
\hat{u}_{{s}} = \hat{\boldsymbol{u}}_{{ns}} \boldsymbol{\cdot} \boldsymbol{t} - \boldsymbol{\delta} \hat{\boldsymbol{u}} \boldsymbol{\cdot} \boldsymbol{t} = \left\{
\begin{array}{ll}
\hat{\beta}_s ~ \hat{\Lambda} ~\left( 1 - \displaystyle\frac{\hat{\tau}_{s}}{\vert \hat{\Lambda} \vert} \right), & \text{iff}~~ \vert \hat{\Lambda} \vert > \hat{\tau}_s, \\[2pt]
0, & \text{iff}~~ \vert \hat{\Lambda} \vert \leqslant \hat{\tau}_s,
\end{array} \right.
\end{equation}
on $\partial X$, where $\hat{u}_{{s}}$ is the slip velocity, $\hat{\boldsymbol{u}}_{{ns}}$ the velocity of the solid boundary $\partial X$, and $\boldsymbol{\delta} \hat{\boldsymbol{u}}$ is the restriction of $\hat{\boldsymbol{u}}$ on $\partial X$: $\hat{\boldsymbol{u}} \to \boldsymbol{\delta} \hat{\boldsymbol{u}}$ as $\hat{\boldsymbol{x}} \to \partial X$. The value of the tangential {\it traction vector} (i.e. tangential force per unit area on the solid surface) is shown by $\hat{\Lambda} = \left[ \left( - \hat{p} \boldsymbol{1} + \hat{\boldsymbol{\tau}} \right) \boldsymbol{\cdot} \boldsymbol{n} \right] \boldsymbol{\cdot} \boldsymbol{t}$, where the normal and tangential unit vectors to the solid surface $\partial X$ are represented by $\boldsymbol{n}$ and $\boldsymbol{t}$, respectively. The no-penetration condition on $\partial X$ reads $\boldsymbol{\delta}\hat{\boldsymbol{u}} \boldsymbol{\cdot} \boldsymbol{n} = \hat{\boldsymbol{u}}_{{ns}} \boldsymbol{\cdot} \boldsymbol{n} $.

Without loss of generality, we fixed $k$ at unity in the slip law (\ref{SlidingLaw}) and will do so in the rest of the present study. This is supported by experimental observations as well: \cite{seth2012soft} validated the slip power index of unity for emulsions contacting non-adhering surfaces. Moreover, in separate experimental studies, \cite{poumaere2014unsteady} and \cite{daneshi2019characterising} reported $k\approx1$ for the pressure-driven flows of Carbopol gels with different concentrations and stirring rates during the sample preparation. However, experiments have also revealed that $k \neq 1$ in some cases \citep{piau2007carbopol,medina2017tangential,medina2019rheo}. We will discuss these cases in Appendix \ref{sec:kneq1}, as the main objectives/conclusions of the present study are independent of the value of $k$.

\subsection{Numerical algorithm}

The minimisation principle for the sliding problem $\mathbb{P}$ can be derived as:
\begin{eqnarray}
\mathcal{J}(\boldsymbol{\hat{u}}) & = & \frac{\hat{\mu}}{2} \int_{\Omega \setminus \bar{X}} \hat{\dot{\boldsymbol{\gamma}}} (\hat{\boldsymbol{u}}) \boldsymbol{:} \hat{\dot{\boldsymbol{\gamma}}} (\hat{\boldsymbol{u}}) ~\text{d}\hat{V} + \hat{\tau}_y \int_{\Omega \setminus \bar{X}} \Vert \hat{\dot{\boldsymbol{\gamma}}} (\hat{\boldsymbol{u}}) \Vert ~\text{d}\hat{V}  - \int_{\Omega \setminus \bar{X}} \hat{\rho} \hat{\boldsymbol{f}} \boldsymbol{\cdot} \hat{\boldsymbol{u}} ~\text{d} \hat{V}\nonumber \\
& + & \frac{1}{2 \hat{\beta}_s} \int_{\partial X} \hat{u}_s^2 ~\text{d}\hat{A} + \hat{\tau}_s \int_{\partial X} \vert \hat{u}_s \vert ~\text{d}\hat{A}.
\end{eqnarray}
For the augmented Lagrangian approach we require to relax the rate of strain tensor to the auxiliary tensor $\hat{\boldsymbol{q}}$ and the restriction of the velocity vector on $\partial X$ to the auxiliary vector $\hat{\boldsymbol{\xi}}$. Then the saddle-point problem associated with $\mathbb{P}$ is:
\begin{eqnarray}
\mathcal{J}(\hat{\boldsymbol{u}},\hat{\boldsymbol{q}},\hat{\boldsymbol{\xi}}) & = & \frac{\hat{\mu}}{2} \int_{\Omega \setminus \bar{X}} \hat{\boldsymbol{q}} \boldsymbol{:} \hat{\boldsymbol{q}} ~\text{d}\hat{V} + \hat{\tau}_y \int_{\Omega \setminus \bar{X}} \Vert \hat{\boldsymbol{q}} \Vert ~\text{d}\hat{V} + \frac{1}{2} \int_{\Omega \setminus \bar{X}} \left[ \hat{\dot{\boldsymbol{\gamma}}} \left( \hat{\boldsymbol{u}} \right) - \hat{\boldsymbol{q}} \right] \boldsymbol{:} \hat{\boldsymbol{T}} ~\text{d}\hat{V} \nonumber \\
& - & \int_{\Omega \setminus \bar{X}} \hat{\rho} \hat{\boldsymbol{f}} \boldsymbol{\cdot} \hat{\boldsymbol{u}} ~\text{d} \hat{V} + \frac{1}{2 \hat{\beta}_s} \int_{\partial X} \hat{\boldsymbol{\xi}}^2 \text{d}\hat{A} + \hat{\tau}_s \int_{\partial X} \vert \hat{\boldsymbol{\xi}} \vert ~\text{d}\hat{A} + \int_{\partial X} \hat{\boldsymbol{\lambda}} \boldsymbol{\cdot} \left( \hat{\boldsymbol{\xi}} - \boldsymbol{\delta} \hat{\boldsymbol{u}} \right) \text{d}\hat{A} \nonumber \\
& + & \frac{a ~\hat{\mu}}{2} \int_{\Omega \setminus \bar{X}}  \left( \hat{\dot{\boldsymbol{\gamma}}} \left( \hat{\boldsymbol{u}} \right) - \hat{\boldsymbol{q}} \right) \boldsymbol{:} \left( \hat{\dot{\boldsymbol{\gamma}}} \left( \hat{\boldsymbol{u}} \right) - \hat{\boldsymbol{q}} \right) \text{d}\hat{V} + \frac{b}{2 \hat{\beta}_s} \int_{\partial X} \left( \hat{\boldsymbol{\xi}} - \boldsymbol{\delta} \hat{\boldsymbol{u}} \right)^2 \text{d}\hat{A},
\end{eqnarray}
where $a$ and $b$ are arbitrary constants (augmentation parameters); $\hat{\boldsymbol{T}}$ and $\hat{\boldsymbol{\lambda}}$ are the Lagrange multipliers.

Hence, the Uzawa algorithm in the presence of slip takes the form of algorithm \ref{alg2}. Upon convergence of the algorithm \ref{alg2} with the free augmentation parameters $a$ and $b$, the Lagrange multiplier $\hat{\boldsymbol{T}}$ converges to the {\it true} stress field, $\hat{\boldsymbol{q}}$ to the {\it true} rate of strain tensor, Lagrange multiplier $\hat{\boldsymbol{\lambda}}$ to the traction vector on $\partial X$, and auxiliary variable $\hat{\boldsymbol{\xi}}$ to the velocity on $\partial X$.

\begin{algorithm}
\caption{}\label{alg2}
\begin{algorithmic}[1]
\Procedure{}{solving yield-stress fluid flow with slip yield stress boundary condition}
\State $m \gets 0$
\vspace{3pt}
\State $ \hat{\boldsymbol{q}}^0,\hat{\boldsymbol{T}}^0,\hat{\boldsymbol{\lambda}}^0,\hat{\boldsymbol{\xi}}^0 \gets \boldsymbol{0}~\text{(or any other initial guess)}$
\State \hspace{-12pt}{\it loop (Uzawa algorithm)}:
\If {$\text{residual} < \text{convergence}$} \textbf{close}.
\EndIf
\State find $\hat{\boldsymbol{u}}^{m+1}$ and $\hat{p}^{m+1}$ which satisfy,

\hspace{-25 pt}
$
\left\{
\begin{array}{l}
-a ~\hat{\mu} ~\boldsymbol{\Delta} \hat{\boldsymbol{u}}^{m+1} = - \boldsymbol{\nabla} \hat{p}^{m+1} + \boldsymbol{\nabla} \boldsymbol{\cdot} \left( \hat{\boldsymbol{T}^m} - a~\hat{\mu} ~\hat{\boldsymbol{q}}^m \right) + \hat{\rho} \hat{\boldsymbol{f}}, ~~\text{in}~ \Omega \setminus \bar{X}, \vspace{8pt} \\[2pt]
a ~\hat{\mu} ~\hat{\dot{\boldsymbol{\gamma}}} (\hat{\boldsymbol{u}}^{m+1}) \boldsymbol{\cdot} \boldsymbol{n} + \displaystyle\frac{b}{\hat{\beta}_s} \boldsymbol{\delta} \hat{\boldsymbol{u}}^{m+1} = a ~\hat{\mu} ~ \hat{\boldsymbol{q}}^m \boldsymbol{\cdot} \boldsymbol{n} + \displaystyle\frac{b}{\hat{\beta}_s}\hat{\boldsymbol{\xi}}^m + \left[ \hat{\boldsymbol{\lambda}}^m - \left( -\hat{p}^{m+1} \boldsymbol{1} + \hat{\boldsymbol{T}^m } \right) \boldsymbol{\cdot} \boldsymbol{n} \right]~\text{on}~\partial X, \vspace{8pt} \\[2pt]
\boldsymbol{\nabla} \boldsymbol{\cdot} \hat{\boldsymbol{u}}^{m+1} = 0, ~~\text{in}~ \Omega \setminus \bar{X},
\end{array} \right.
$

with given B.C.: $\hat{\boldsymbol{u}}^{m+1}=\hat{\boldsymbol{u}}_0$~~on $\partial \Omega$.
\vspace{5 pt}
\State $\hat{\boldsymbol{q}}^{m+1} \gets \left\{
\begin{array}{ll}
~~~~~~~~~~~~~0, & \text{iff}~~ \Vert \hat{\boldsymbol{\Sigma}} \Vert \leqslant \hat{\tau}_y, \\[2pt]
\left( 1- \displaystyle\frac{\hat{\tau}_y}{\Vert\hat{\boldsymbol{\Sigma}}\Vert} \right) \displaystyle\frac{\hat{\boldsymbol{\Sigma}}}{(1+a) \hat{\mu}}, & \text{iff}~~ \Vert \hat{\boldsymbol{\Sigma}} \Vert > \hat{\tau}_y.
\end{array} \right.
$

\vspace{5pt}
where $\hat{\boldsymbol{\Sigma}} = \hat{\boldsymbol{T}}^m+a~\hat{\mu} ~\hat{\dot{\boldsymbol{\gamma}}} \left(\hat{\boldsymbol{u}}^{m+1}\right)$.
\vspace{5 pt}

\vspace{5pt}
\State $\hat{\boldsymbol{\xi}}^{m+1} \gets \left\{
\begin{array}{ll}
~~~~~~~~~~\left( \hat{\boldsymbol{u}}_{{ns}} \boldsymbol{\cdot} \boldsymbol{n} \right) \boldsymbol{n} + \left( \hat{\boldsymbol{u}}_{{ns}} \boldsymbol{\cdot} \boldsymbol{t} \right) \boldsymbol{t} , & \text{iff}~~ \vert \hat{\Phi} \vert \leqslant \hat{\tau}_s, \\[2pt]
\displaystyle \left( \hat{\boldsymbol{u}}_{{ns}} \boldsymbol{\cdot} \boldsymbol{n} \right) \boldsymbol{n} + \left( \hat{\boldsymbol{u}}_{{ns}} \boldsymbol{\cdot} \boldsymbol{t} \right) \boldsymbol{t} + \frac{\hat{\beta}_s}{1+ b} ~\hat{\Phi} \left( 1 - \frac{\hat{\tau}_s}{\vert \hat{\Phi} \vert} \right)~\boldsymbol{t}, & \text{iff}~~ \vert \hat{\Phi} \vert > \hat{\tau}_s.
\end{array} \right.
$

\vspace{5pt} where $\hat{\Phi} =  - \hat{\boldsymbol{\lambda}}^m \boldsymbol{\cdot} \boldsymbol{t} + \displaystyle \frac{b}{\hat{\beta}_s}~ \left( \boldsymbol{\delta} \hat{\boldsymbol{u}}^{m+1} \boldsymbol{\cdot} \boldsymbol{t} - \hat{\boldsymbol{u}}_{ns} \boldsymbol{\cdot} \boldsymbol{t} \right)  $
\vspace{5pt}

\State $\hat{\boldsymbol{T}}^{m+1} \gets \hat{\boldsymbol{T}}^m + a~\hat{\mu} \left[ \hat{\dot{\boldsymbol{\gamma}}} \left(\hat{\boldsymbol{u}}^{m+1}\right) - \hat{\boldsymbol{q}}^{m+1} \right]$

\vspace{5pt}
\State $\hat{\boldsymbol{\lambda}}^{m+1} \gets \hat{\boldsymbol{\lambda}}^m - \displaystyle \frac{b}{\hat{\beta}_s} \left[ \boldsymbol{\delta} \hat{\boldsymbol{u}}^{m+1} - \hat{\boldsymbol{\xi}}^{m+1} \right]$
\vspace{5pt}

\State residual $\gets$
\begin{eqnarray*}
\max \biggl( \displaystyle \int_{\Omega \setminus \bar{X}} \vert \hat{\boldsymbol{u}}^{m+1} - \hat{\boldsymbol{u}}^m \vert &\text{d}\hat{A}&, \displaystyle \int_{\Omega \setminus \bar{X}} \Vert \hat{\dot{\boldsymbol{\gamma}}} \left( \hat{\boldsymbol{u}}^{m+1}\right) - \hat{\boldsymbol{q}}^{m+1} \Vert ~\text{d}\hat{A} , \nonumber \\
\displaystyle \int_{\Omega \setminus \bar{X}} \Vert \hat{\boldsymbol{q}}^{m+1} - \hat{\boldsymbol{q}}^{m} \Vert ~ &\text{d}\hat{A}&, \int_{\partial X} \vert \hat{\boldsymbol{\lambda}}^{m+1} - \hat{\boldsymbol{\lambda}}^m \vert ~\text{d}\hat{S} \biggl)
\end{eqnarray*}

\State $m \gets m+1$
\State \textbf{goto} \emph{loop}.
\EndProcedure
\end{algorithmic}
\end{algorithm}

A mesh adaptation procedure, the same as the one proposed by \cite{roquet2003adaptive}, could be coupled with the above algorithm to obtain a fine resolution of the yield surfaces. Based on this procedure, the adapted mesh is stretched anisotropically in the direction of the eigenvectors of the Hessian of $\sqrt{\hat{\mu} ~\hat{\dot{\boldsymbol{\gamma}}} \boldsymbol{:} \hat{\dot{\boldsymbol{\gamma}}} + \hat{\tau}_y \Vert \hat{\dot{\boldsymbol{\gamma}}} \Vert}$, which is the square root of the local energy dissipation. In this study, we implement the entire numerical algorithm in an open-source C++ finite element environment---FreeFEM++ \citep{MR3043640}. We have previously validated our numerical implementation and mesh adaptation widely within various studies \citep{chaparian2017yield,chaparian2019adaptive,chaparian2019porous}. Here we will not go into technical details such as the convergence rate, since such details are well-documented in the previous studies; please see for example \cite{roquet2008adaptive}. 

In what follows, we quickly validate the algorithm for the sliding flows and most importantly drive variational tools \citep{mosolov1965variational} for a sample problem. These tools will be used to analyse more complex flows in the next sections.

\section{Benchmark problem: sliding channel Poiseuille flow}
In this section we consider the sliding flow of a Bingham fluid in a channel (Poiseuille flow); the same as the one shown in figure \ref{fig:Schematic_slip}. The walls are denoted by $\Gamma$ and the full domain by $\Omega$. We consider two classic formulations: (i) [M]obility problem where the pressure gradient is applied and the flow rate can be computed and (ii) [R]esistance problem in which flow rate is set and as a result pressure gradient can be computed. We also validate the presented numerical algorithm and derive/revisit some variational tools \citep{huilgol1998variational} useful for the rest of this study.

\subsection{{\normalfont [M]} problem}

In this subsection, we use $\hat{G} \hat{H}^2 /\hat{\mu}$ as the velocity scale, $\hat{G}\hat{H}$ as the characteristic viscous stress to scale the deviatoric stress tensor and $\hat{H} / \hat{\mu}$ to scale the slip coefficient where $\hat{G}$ is the absolute value of the applied pressure-gradient to drive the flow from left to right (in the positive direction of the $x-$axis). Hence, the non-dimensional governing and constitutive equations take the forms:
\begin{equation}
\boldsymbol{e}_x + \boldsymbol{\nabla} \boldsymbol{\cdot} \boldsymbol{\tau} = 0
\end{equation}
and,
\begin{equation}\label{non-const-M}
  \left\{
    \begin{array}{ll}
      \boldsymbol{\tau} = \left( 1 + \displaystyle{\frac{Od}{\Vert\dot{\boldsymbol{\gamma}}\Vert}} \right) \dot{\boldsymbol{\gamma}} & \mbox{iff}\quad \Vert \boldsymbol{\tau} \Vert > Od, \\[2pt]
      \dot{\boldsymbol{\gamma}} = 0 & \mbox{iff}\quad \Vert \boldsymbol{\tau} \Vert \leqslant Od,
  \end{array} \right.
\end{equation}
respectively, and the slip law,
\begin{equation}\label{SlipLaw-M}
u_{s} = \left\{
\begin{array}{ll}
\beta_s ~\Lambda ~\left( 1 - \displaystyle\frac{Od_s}{\vert \Lambda \vert} \right), & \text{iff}~~ \vert \Lambda \vert > Od_s, \\[2pt]
0, & \text{iff}~~ \vert \Lambda \vert \leqslant Od_s,
\end{array} \right.
\end{equation}
where $Od = \frac{\hat{\tau}_y}{\hat{H} \hat{G}}$ and $Od_s = \frac{\hat{\tau}_s}{\hat{H} \hat{G}}$. In what follows, the ratio $\frac{\hat{\tau}_s}{\hat{\tau}_y}$ is denoted by $\alpha_s$. The analytical solutions to the velocity and stress fields are derived in Appendix \ref{app:M}.

The energy balance equation in the presence of slip is:
\begin{eqnarray}
& & a(\boldsymbol{u},\boldsymbol{u}) + Od ~j(\boldsymbol{u}) + \int_{\Gamma} \vert \sigma_{nt} ~u_s \vert ~\text{d}S =  \nonumber \\
& & a(\boldsymbol{u},\boldsymbol{u}) + Od ~j(\boldsymbol{u}) + \frac{1}{\beta_s} \int_{\Gamma}  u^2_s ~\text{d}S + Od_s \int_{\Gamma} \vert u_s \vert ~\text{d}S  = \nonumber \\
& & a(\boldsymbol{u},\boldsymbol{u}) + Od ~j(\boldsymbol{u}) + a_s (u_s,u_s) + Od_s ~j_s(u_s)  = \int_{\Omega} ~ \boldsymbol{u} \boldsymbol{\cdot} \boldsymbol{e}_x ~\text{d}A,
\end{eqnarray}
where $a(\boldsymbol{u},\boldsymbol{u}) = \int_{\Omega} \dot{\boldsymbol{\gamma}} (\boldsymbol{u}) \boldsymbol{:} \dot{\boldsymbol{\gamma}} (\boldsymbol{u}) ~\text{d} A$ and $Od~j(\boldsymbol{u}) = Od~\int_{\Omega} \Vert \dot{\boldsymbol{\gamma}} (\boldsymbol{u}) \Vert ~\text{d} A$
are the viscous and plastic dissipations, respectively. Please note that the slip dissipation $\int_{\partial X} \vert \sigma_{nt} ~u_s \vert ~\text{d}S$ can be split into two terms: the `viscous' slip dissipation $(1/\beta_s) \int_{\Gamma} u_s^2~\text{d}S$ which is designated by $a_s (u_s,u_s)$ in this study and the `plastic' slip dissipation $Od_s \int_{\Gamma} \vert u_s \vert ~\text{d}S$ with $Od_s ~j_s(u_s)$. Moreover, $j(\boldsymbol{u})$ is the normalized plastic dissipation and $j_s(u_s)$ the normalized `plastic' slip dissipation.

We can rearrange the energy balance equation and form a set of inequalities:
\begin{eqnarray}\label{OdcDefDerivation}
0 \leqslant a(\boldsymbol{u},\boldsymbol{u}) = & -j(\boldsymbol{u}) & \left( Od - \frac{\displaystyle \int_{\Omega} ~ \boldsymbol{u} \boldsymbol{\cdot} \boldsymbol{e}_x ~\text{d}A - a_s (u_s,u_s) - Od_s ~j_s (u_s)}{j(\boldsymbol{u})} \right)  \leqslant \nonumber \\
& -j(\boldsymbol{u}) & \left( Od - \frac{\displaystyle \int_{\Omega} ~ \boldsymbol{u} \boldsymbol{\cdot} \boldsymbol{e}_x ~\text{d}A - Od_s ~j_s (u_s)}{j(\boldsymbol{u})} \right) \leqslant  \nonumber \\
& -j(\boldsymbol{u}) & \left( Od - \sup_{\boldsymbol{v} \in \boldsymbol{V}, ~\boldsymbol{v} \neq \boldsymbol{0}} \frac{\displaystyle \int_{\Omega} ~ \boldsymbol{v} \boldsymbol{\cdot} \boldsymbol{e}_x ~\text{d}A - Od_s ~j_s (v_s)}{j(\boldsymbol{v})} \right),
\end{eqnarray}
to find the definition of the critical Oldroyd number in the presence of slip as,
\begin{equation}\label{eq:OdcDef}
Od_c =  \sup_{\boldsymbol{v} \in \boldsymbol{V}, ~\boldsymbol{v} \neq \boldsymbol{0}} \left( \frac{\displaystyle \int_{\Omega} ~ \boldsymbol{v} \boldsymbol{\cdot} \boldsymbol{e}_x ~\text{d}A - Od_s ~j_s (v_s)}{j(\boldsymbol{v})} \right),
\end{equation}
where $\boldsymbol{v}$ is a velocity test function from the set of all admissible velocity fields $\boldsymbol{V}$. Then it enforces that for $Od_c \leqslant Od$, $a(\boldsymbol{u},\boldsymbol{u})=0$ or $\boldsymbol{u}=0$ (i.e. no flow condition).

We may use (\ref{eq:OdcDef}) for calculating $Od_c$ in simple flows such as Poiseuille flow here: the flow can be postulated by two boundary layers with thickness $\delta$ attached to the walls, through which the slip velocity $U_1$ is connected to the plug velocity $U_2$, then,
\[
j(\boldsymbol{v}) \approx 2\frac{U_2 - U_1}{\delta} \delta \ell = 2(U_2 - U_1) \ell~~\&~~ j_s (v_s) = 2 U_1 \ell,
\]
where $\ell$ is the length of the chosen control volume. This suggests,
\[
Od_c = \sup_{0<U_1 \leqslant U_2, ~(U_1,U_2) \in \mathbb{R}} \frac{U_2 - 2 ~Od_s ~U_1 }{2 (U_2 - U_1)} = \sup_{0<U_1 \leqslant U_2, ~(U_1,U_2) \in \mathbb{R}} \frac{U_2 - 2 \alpha_s ~Od_c ~U_1 }{2 (U_2 - U_1)},
\]
or,
\begin{equation}
Od_c = \sup_{0<U_1 \leqslant U_2, ~(U_1,U_2) \in \mathbb{R}}  \frac{1}{2 [1 + (\alpha_s-1) U_1 / U_2]}.
\end{equation}
The maximal value of the argument is $1/2\alpha_s$ which occurs at $U_1 / U_2 = 1$. Its physical interpretation is that the supremum occurs in the fully sliding regime which is intuitive. Hence, $Od_c = 1/2\alpha_s$ or in the other words, there is no flow if $1/2 \leqslant Od_s$, which is in agreement with the solution of the full equations derived in \S \ref{app:M}.

\begin{figure}
\centerline{\includegraphics[width=0.8\linewidth]{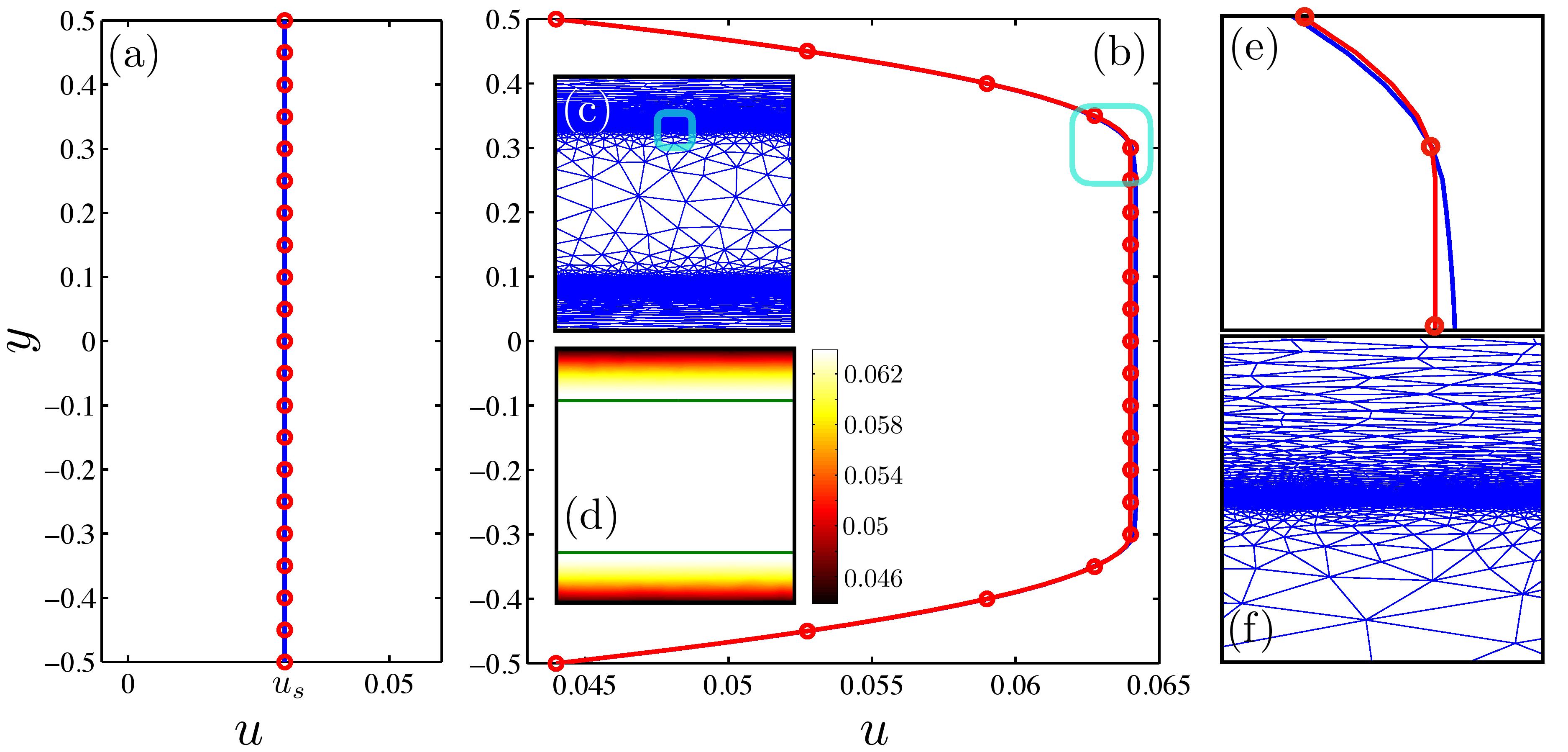}}
\caption{[M] problem features: (a) velocity profile of the case $Od=1, \alpha_s=0.2, \beta_s=0.1$ (fully sliding plug), (b) velocity profile of the case $Od=0.3, \alpha_s=0.2, \beta_s=0.1$ (deforming regime), (c) the mesh after 5 cycles of adaptation associated with panel (b), ~(d) Velocity contour associated with panel (b); the green lines show the yield surfaces, (e,f) zoom in of the cyan windows in the panels (b,c) respectively. Please note that in panels (a,b,e) the blue line shows the numerical data corresponding to the initial mesh, the red line corresponds to the adapted mesh (the mesh shown in the panel (c)), and the red circles are the analytical solution; see Appendix \ref{app:M}.}
\label{fig:ValidationM}
\end{figure}

We perform a few simulations for the [M] problem to benchmark the presented algorithm. Figure \ref{fig:ValidationM} illustrates different regimes: panel (a) compares the computed velocity profile (blue line) in the gap with the analytical solution (red circles) in the fully sliding plug regime ($Od=1, \alpha_s=0.2 ~\&~ \beta_s=0.1$) and panel (b) within the deforming regime. Utilizing the mesh adaptation seems more indispensable in the deforming regime compared to the fully sliding plug regime where velocity gradient is zero. It is clear that the velocity distribution after adaptation (red line) is much closer to the analytical solution (red circles) than the solution of the initial mesh (blue line), and also yields a fine resolution of the yield surfaces; see panel (d). We shall mention that the focus of the present study is not on quantifying these improvements by the mesh adaptation as it has been previously studied in details \citep{roquet2003adaptive,roquet2008adaptive,chaparian2019adaptive}.

\subsection{{\normalfont [R]} problem}\label{sec:R}
Here, we reformulate the problem in the resistance scalings and the quantities are designated with asterisk sign ($*$) to prevent any potential confusion with the [M] problem. In the resistance formulation, the velocity is scaled with the mean bulk velocity, $\hat{U}=\hat{Q}/\hat{H}$, and as a result, there is always a non-zero flow with the flow rate equals to unity ($Q^*=1$), whether the flow is in the deforming regime or the fully sliding plug regime. The deviatoric stress and pressure are scaled with the characteristic viscous stress, $\hat{\mu} \hat{U}/\hat{H}$, hence,
\begin{equation}\label{eq:Resistance}
G^* \boldsymbol{e}_x + \boldsymbol{\nabla} \boldsymbol{\cdot} \boldsymbol{\tau}^* = 0,
\end{equation}
and,
\begin{equation}\label{non-const-R}
  \left\{
    \begin{array}{ll}
      \boldsymbol{\tau}^* = \left( 1 + \displaystyle{\frac{B}{\Vert\dot{\boldsymbol{\gamma}}^*\Vert}} \right) \dot{\boldsymbol{\gamma}}^* & \mbox{iff}\quad \Vert \boldsymbol{\tau}^* \Vert > B, \\[2pt]
      \dot{\boldsymbol{\gamma}}^* = 0 & \mbox{iff}\quad \Vert \boldsymbol{\tau}^* \Vert \leqslant B,
  \end{array} \right.
\end{equation}
with the boundary condition,
\begin{equation}\label{SlipLaw-R}
u^*_{{s}} = \left\{
\begin{array}{ll}
\beta_s ~\Lambda^* ~\left( 1 - \displaystyle\frac{B_{s}}{\vert \Lambda^* \vert} \right), & \text{iff}~~ \vert \Lambda^* \vert > B_{s}, \\[2pt]
0, & \text{iff}~~ \vert \Lambda^* \vert \leqslant B_{s},
\end{array} \right.
\end{equation}
on the walls govern the [R] problem. Please note that $G^*$ is the absolute value of the pressure gradient which enforces the unity flow rate for a given Bingham number $\left( \frac{\hat{\tau}_y \hat{H}}{\hat{\mu} \hat{U}} \right)$, $\alpha_s$ and $\beta_s$. The energy balance equation in this case is,
\begin{eqnarray}\label{energyR}
& & a(\boldsymbol{u}^*,\boldsymbol{u}^*) + B~ j(\boldsymbol{u}^*) + \int_{\Gamma} \vert \sigma^*_{nt} ~u^*_s \vert ~\text{d}S = \nonumber \\
& &  a(\boldsymbol{u}^*,\boldsymbol{u}^*) + B~ j(\boldsymbol{u}^*) + \frac{1}{\beta_s} \int_{\Gamma} {u^*_s}^2 ~\text{d}S + B_s \int_{\Gamma} \vert u^*_s \vert ~\text{d}S = \nonumber \\
& &  a(\boldsymbol{u}^*,\boldsymbol{u}^*) + B~ j(\boldsymbol{u}^*) + a_s (u^*_s,u^*_s) + B_s ~j_s (u^*_s) = \nonumber \\
& & \int_{\Omega} \frac{\text{d} p^*}{\text{d} x}~ \boldsymbol{u}^* \boldsymbol{\cdot} \boldsymbol{e}_x ~\text{d}A = G^* \int_{\Omega} ~ \boldsymbol{u}^* \boldsymbol{\cdot} \boldsymbol{e}_x ~\text{d}A.
\end{eqnarray}

\cite{frigaard2019background} has shown that at the yield limit ($B \to \infty$), viscous dissipation is at least one order of magnitude less than the plastic dissipation (when $B \to \infty$: $a(\boldsymbol{u}^*,\boldsymbol{u}^*) \ll B~ j(\boldsymbol{u}^*)$). With a little extra effort, we can show that this applies to $a_s (u^*_s,u^*_s)$ and $B_s ~j_s (u^*_s)$ as well, knowing that $u_s^* \leqslant 1$ (from the continuity equation $Q^*=1$) and $B_s \to \infty$ in this limit. Exploiting that generally the mapping between [M] and [R] problems is attainable via $Q = \frac{Od}{B}$ or $G^* = \frac{B}{Od}$, one can extract the critical Oldroyd number from the [R] formulation through,

\begin{equation}\label{OdcBtoInf}
Od_c = \lim_{B \to \infty} \frac{\displaystyle \int_{\Omega} ~ \boldsymbol{u}^* \boldsymbol{\cdot} \boldsymbol{e}_x ~\text{d}A}{j(\boldsymbol{u}^*)+\displaystyle \frac{1}{B} \int_{\Gamma} \vert \sigma^*_{nt} ~u^*_s \vert ~\text{d}S} = \displaystyle \lim_{B \to \infty} \frac{HL}{j(\boldsymbol{u}^*)+\alpha_s ~j_s (u^*_s)}.
\end{equation}

\begin{figure}
\centerline{\includegraphics[width=0.5\linewidth]{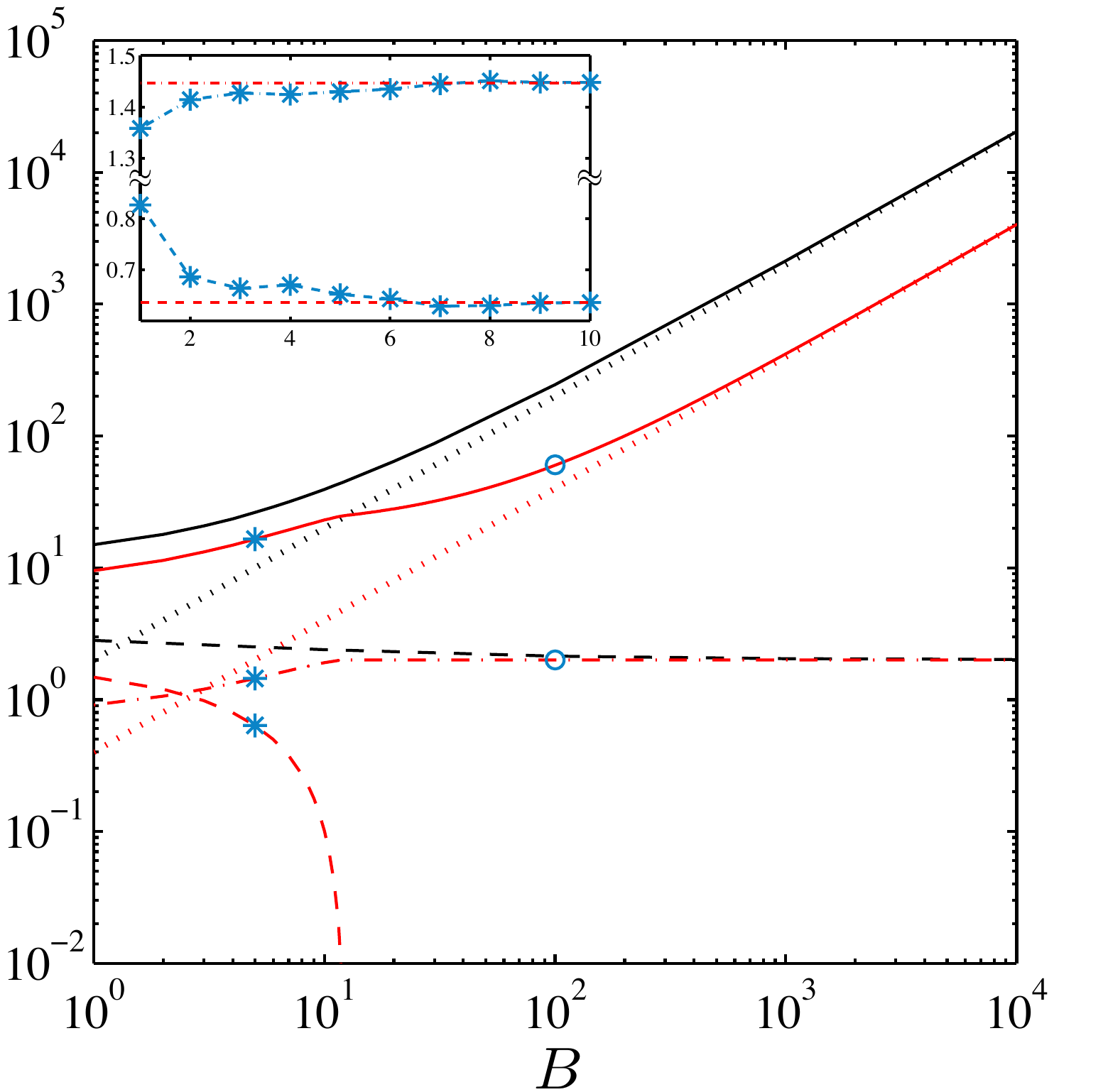}}
\caption{Features of the [R] problem for no-slip (in black) and sliding flow $\alpha_s=0.2 ~\&~ \beta_s = 0.1$ (in red). Lines represent analytically calculated values (see Appendix \ref{app:R}): continuous lines show  $G^*$, dashed lines $j(\boldsymbol{u}^*)$, and dashed-dotted line $j_s (u_s)$. The black dotted line is the asymptotic scaling $2 B$ and red dotted line $2 \alpha_s B = 0.4 B$. The asterisk symbols mark the computational data in the deforming regime and circle symbols in the fully-sliding regime. Inset ($B=5, B_s = 1 ~\&~ \beta_s =0.1$): $j(\boldsymbol{u}^*)$ (cyan dashed line and asterisks) and $j_s (u_s)$ (cyan dashed-dotted line and asterisks) versus the cycles of mesh adaptation index with red lines showing the analytically calculated values in Appendix \ref{app:R} .}
\label{fig:Resistance}
\end{figure}

To explain the relation between [R] and [M] problems further, we consider some examples in figure \ref{fig:Resistance}. The analytical solutions to equations (\ref{eq:Resistance}-\ref{SlipLaw-R}) of the [R] problem are derived in Appendix \ref{app:R}. The absolute value of the pressure gradient versus the Bingham number under the no-slip condition and the sliding flow $\alpha_s=0.2~\&~\beta_s=0.1$ are shown with the continuous black and red lines, respectively. While for both type of flows an increase in $G^*$ by increasing Bingham number can be observed, they display different trends. This can be explained by knowing that beyond a critical Bingham number ($\bar{B} = 1/\beta_s (1-\alpha_s) \leqslant B$), the sliding flow state switches from the deforming regime to the fully-sliding plug regime. Only at large enough Bingham numbers ($B \to \infty$), we can use the numerical data to extract the yield limit. This is illustrated in figure \ref{fig:Resistance} by the dotted lines, where at $B \to \infty$, $G^*$ asymptotes to $2B$ and $2 \alpha_s B = 0.4 B$ for no-slip and the sample sliding flow considered, respectively. In this limit, $j(\boldsymbol{u}^*)$ (dashed lines in figure \ref{fig:Resistance}) asymptotes to 2 and 0 for the no-slip and sliding flows: this is equivalent to $Od_c = 1/2$ and $1/2\alpha_s$, respectively, knowing that $j_s (u_s)$ asymptotes to 2 for the sliding flows at sufficiently large Bingham numbers (follow the red dashed-dotted line in figure \ref{fig:Resistance}). These results are the same as the ones derived in Appendix \ref{app:M} considering the [M] formulation. Please note that due to the log-log scaling used in figure \ref{fig:Resistance}, $j(\boldsymbol{u}^*)$ corresponding to the sliding flow, which is exactly equal to zero for $\bar{B} \leqslant B$ is not visible; however, its fast decay is clear: $j(\boldsymbol{u}^*)$ is almost equal to $10^{-2}$ for $B\approx11.85$ where $\bar{B}(\alpha_s=0.2~\&~\beta_s=0.1) = 12.5$. From a different perspective, figure \ref{fig:Resistance} inset shows how mesh adaptation helps us to achive more accurate results.

\section{Slippery particle motion}
Particle motion in yield-stress fluids has been considered in numerous studies ranging from numerical simulations \citep{beris1985creeping,tokpavi2008very,putz2010creeping,nirmalkar2012creeping,chaparian2017cloaking,chaparian2017yield,fraggedakis2016soft,izbassarov2018computational} to experimental validations/extensions \citep{jossic2001,putz2008settling,tokpavi2009experimental}. Unfortunately, less studies are devoted to investigate the effect of slip on the flow field around the particle surface. Although \cite{fraggedakis2016soft} used Navier-slip law in their simulations, the emphasis of their study was on the effect of elasticity of the yield-stress fluid in the particle sedimentation problem. On the experimental side, \cite{jossic2001} reported the drag coefficient of the particle moving in a Carbopol gel and showed that for particles with smooth surfaces, the drag coefficient is smaller compared to the rough particles that prevent slip which is intuitive. However, the mentioned authors did not quantify the slip velocity/law on the particle surface and rather qualitatively addressed the effect of slip. In this section, we will shed light on the effect of slippery motion of the yield-stress fluid about a circular 2D particle and in details will address the yield limit under this condition. 

Flow configuration: the horizontal symmetry axis of the particle is aligned with the $x-$axis (positive direction to the right) and the vertical one by $y-$axis (positive direction upward) where the origin of the coordinate system is fixed at the particle centre. The fluid flow about the particle ($X$) with its radius $\hat{R}$ as the length scale can be described again by two formulations:
\begin{itemize}
\item[(i)] [R] formulation, where the particle is dragged through the fluid with a prescribed velocity $\hat{V}_p^*$ as the boundary condition. Hence, the drag force $\hat{F}^*_D$ can be calculated. In this formulation, $\hat{V}_p^*$ is used as the velocity scale and $\hat{\mu} \hat{V}_p^* / \hat{R}$ as the characteristic viscous stress to scale the deviatoric stress tensor and the pressure, hence the governing equation is,
\begin{equation}
-\boldsymbol{\nabla}p^* + \boldsymbol{\nabla} \boldsymbol{\cdot} \boldsymbol{\tau}^*=0~~\text{in} ~\Omega \setminus \bar{X}~~\&~~\boldsymbol{u}_{ns}^*=-\boldsymbol{e}_y~~\text{on}~\partial X
\end{equation}
with (\ref{non-const-R}) and (\ref{SlipLaw-R}) as the constitutive and the slip law, respectively, where $B=\hat{\tau}_y \hat{R}/ \hat{\mu} \hat{V}^*_p$. The drag force on the particle can be computed as a result from,
\begin{equation}
a(\boldsymbol{u}^*,\boldsymbol{u}^*) + B j(\boldsymbol{u}^*) + a_s (u^*_s,u^*_s) + B_s ~j_s (u^*_s) = \int_{\partial X} \left( \boldsymbol{\sigma}^* \boldsymbol{\cdot} \boldsymbol{n} \right) \boldsymbol{\cdot} \boldsymbol{e}_y ~\text{d}S = F^*_D.
\end{equation}
\item[(ii)] [M] formulation, where the buoyancy of the particle is known and it falls freely under the action of the gravity $-\hat{g} ~\boldsymbol{e}_y$. Hence, the settling velocity of the particle $\hat{V}_p$ can be calculated as a result. In this formulation, velocity is scaled with $\Delta \hat{\rho} \hat{g} \hat{R}^2 / \hat{\mu}$: the velocity scale obtained via balancing the buoyancy stress with the characteristic viscous stress. Please note that here $\hat{g}$ is the gravitational acceleration and $\Delta \hat{\rho}$ is the density difference between the particle and the suspending fluid. Hence, the governing equation is,
\begin{equation}
-\boldsymbol{\nabla}p + \boldsymbol{\nabla} \boldsymbol{\cdot} \boldsymbol{\tau} - \frac{\rho}{1-\rho} \boldsymbol{e}_y =0~~\text{in} ~\Omega \setminus \bar{X}~~\&~~\int_{\partial X} \left( \boldsymbol{\sigma} \boldsymbol{\cdot} \boldsymbol{n} \right) \boldsymbol{\cdot} \boldsymbol{e}_y ~\text{d}S=\frac{\pi}{1-\rho}
\end{equation}
where $\rho$ is the ratio of the fluid density to the particle one (i.e.~$\hat{\rho}_f / \hat{\rho}_p$). The constitutive and slip laws are the same as (\ref{non-const-M}) and (\ref{SlipLaw-M}), respectively, with a slight difference: the yield number $Y$ is substituted for the Oldroyd number in the particle sedimentation problem. Indeed, $Y=\hat{\tau}_y / \Delta\hat{\rho} \hat{g} \hat{R}$.
\end{itemize}

The yield limit can be well understood using either [R] or [M] formulations knowing the mapping between these two formulations: $Y = \pi B / F^*_D$ or $B=Y/V_p$.  In [R] formulation, the particle never stops because the particle velocity is set as the boundary condition, yet, in the yield limit $B \to \infty$, the drag force also goes to infinity. However, in the [M] problem, for a large enough yield number $(Y > Y_c)$, the yield stress can overcome the buoyancy of the particle and makes it stationary suspended. The connection between these two situations, can be well discussed by the `plastic drag coefficient',
\begin{equation}
  C_D^p = \left\{
    \begin{array}{ll}
      \left[ \frac{\hat{F}^*_D}{\hat{\ell}_\bot \hat{\tau}_Y} \right]^{[R]} = \left[ \frac{F^*_D}{\ell_\bot B} \right]^{[R]} & \mbox{for problem [R]}, \\[2pt]
      \left[ \frac{ \Delta\hat{\rho} \hat{g} \hat{A}_p}{\hat{\ell}_\bot \hat{\tau}_Y} \right]^{[M]} = \left[ \frac{\pi}{\ell_\bot Y} \right]^{[M]} & \mbox{for problem [M]},
  \end{array} \right.
\end{equation}
where $\ell_\bot$ is the width of the particle perpendicular to the flow direction ($\ell_\bot=2$) and $\hat{A}_p$ is the area of the 2D particle (i.e.~$\pi \hat{R}^2$ here). Then, the critical plastic drag coefficient (i.e.~plastic drag coefficient at the yield limit) can be converted easily to the critical yield number,
\begin{equation}
C_{D,c}^p = \left[ C_{D,c}^p \right]^{[R]}_{B \to \infty} = \left[ C_{D,c}^p \right]^{[M]}_{Y \to Y_c^-} = \frac{\pi}{\ell_\bot Y_c}.
\end{equation}
For more details please see \cite{chaparian2017yield}. For instance, for a 2D circle, as has been extensively validated \citep{tokpavi2008very,putz2010creeping,chaparian2017cloaking,chaparian2017yield}, the critical plastic drag coefficient under the no-slip condition is 11.94 which yields to $Y_c=0.1316$.

Regarding the computational details, we conduct the same strategy as in our previous studies \citep{chaparian2017cloaking,chaparian2017yield,iglesias2020computing}. The computational box ($\Omega$) is chosen large enough to ensure independency from far-field conditions \citep{chaparian2018inline}. The boundary condition $\boldsymbol{u}^*=0$ is enforced on $\partial \Omega$.

The yield limit of particle motion in a quiescent yield-stress fluid is directly relevant to the lateral resistance of an object (\eg a pile) in a perfectly-plastic medium (\eg soil) which can be investigated by method of characteristics (i.e. slipline solution) due to the hyperbolic nature of the equations. We briefly revisit this problem in the next subsection.

\subsection{Revisit slipline solution and lower \& upper bound calculations}

In a deformed perfectly-plastic medium, the second invariant of the stress is equal to $B$ everywhere. The equilibrium equations for the unrestricted 2D plastic flow then form a closed set of hyperbolic equations for which there are two family of orthogonal characteristic lines: $\alpha-$ and $\beta-$family \citep{hill1998mathematical,chakrabarty2012theory}. Physically, these lines (termed as sliplines) show the direction of the maximum shear stress which is equal to $B$. Properties of these lines facilitate studying the `yield limit' in various viscoplastic problems \citep{chaparian2017cloaking,chaparian2017yield,dubash2009final,chaparian2018box,hewitt2018viscoplastic}. However, finding the sliplines itself is not trivial in some complex problems. Indeed, the sliplines assist us in postulating admissible stress and velocity fields which yield to finding the lower and upper bounds, respectively, of the load limit.

Initially investigated by \cite{randolph1984limiting}, a slipline solution was devised for calculating the lateral resistance of a circular pile in soil (considered as a perfectly-plastic material), taking into account the effect of soil adhesion at the pile-soil interface, $\tilde{\alpha_s}$. In other words, the tangential shear stress at the pile-soil interface is assumed to be equal to $\tilde{\alpha}_s B$. Although widely believed that the \cite{randolph1984limiting} solution is exact (lower and upper bounds of the load/drag are the same for the whole range of $\tilde{\alpha}_s$), an issue was firstly detected by \cite{murff1989pipe}: for $\tilde{\alpha}_s < 1$, there is a region in the vicinity of the pile surfaces in which the rate of strain is negative and its absolute value should be taken into account in calculating the upper bound to avoid negative plastic dissipation. If one does so, then there is a discrepancy between the lower and upper bounds. Later, \cite{martin2006upper} proposed two new postulated velocity fields which improved the upper bound predictions to some extent; one for small ($\tilde{\alpha}_s \to 0$) and the other one for large ($\tilde{\alpha}_s \to 1$) soil adhesion factors. By combining these two mechanisms, \cite{martin2006upper} proposed an alternative mechanism which markedly shrinks the uncertainty between the lower and upper bounds predictions for the entire range of $\tilde{\alpha}_s$. We will quickly review the lower \citep{randolph1984limiting} and upper \citep{martin2006upper} bound solutions here; however, for more details readers are refereed to the original references.

\begin{figure}
\centerline{\includegraphics[width=0.5\linewidth]{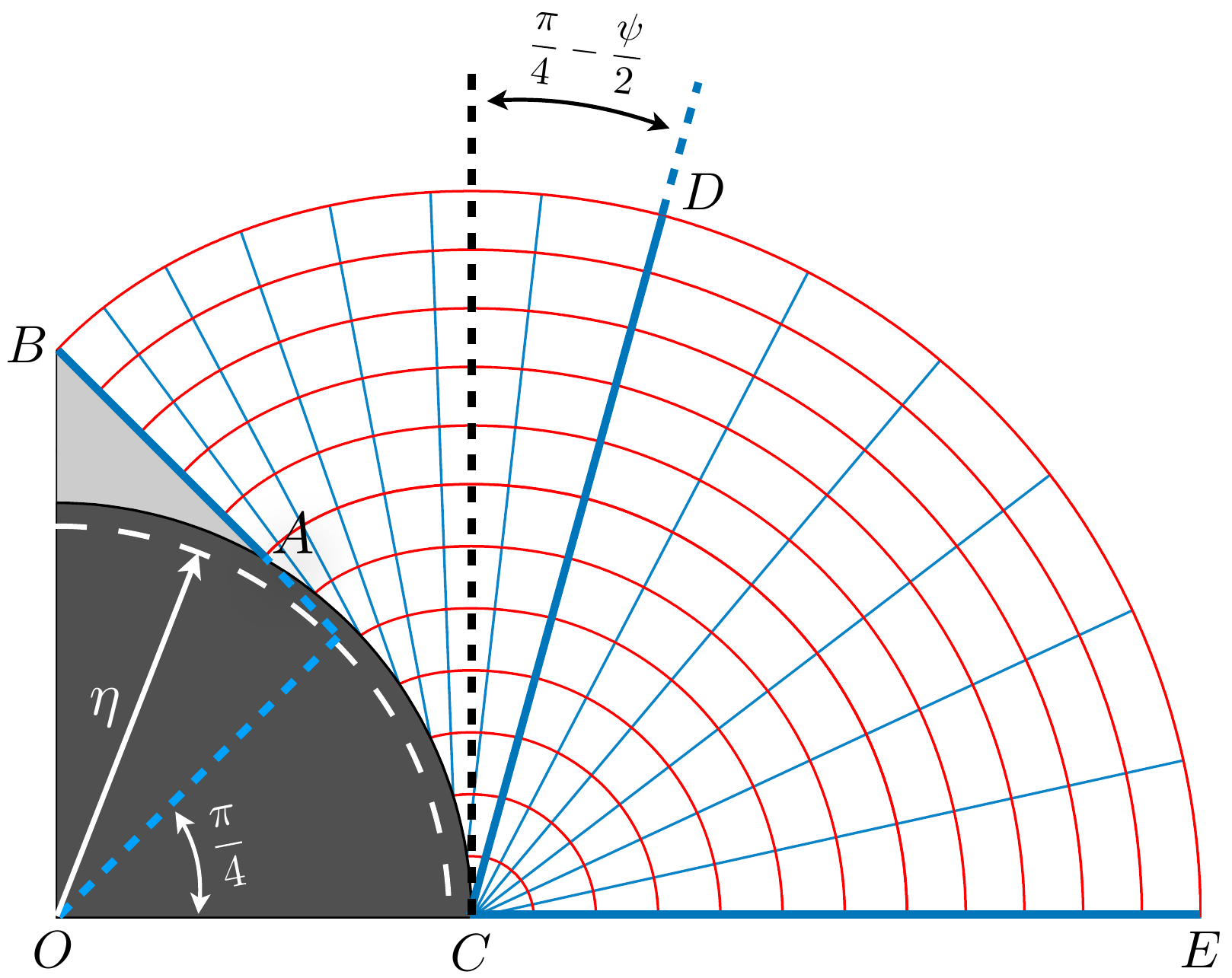}}
\caption{Schematic of the sliplines about a 2D circle. $\alpha-$lines are painted in cyan colour and $\beta-$lines in red. Only a quarter of the whole domain is shown due to symmetry.}
\label{fig:schematic_sliplines}
\end{figure}

The first family of sliplines in this problem are straight lines that make $\frac{\pi}{4}-\frac{\psi}{2}$ angle with the pile surface (say $\alpha-$lines) where $\psi=\sin^{-1} \tilde{\alpha}_s$. Figure \ref{fig:schematic_sliplines} represents these lines in cyan colour. The initial $\alpha-$line is $AB$ which makes $\pi/4$ angle with the vertical symmetry line since $\tilde{\tau}_{xy}$ is zero on $OB$. Hence, the region enclosed between $AB$ and the pile surface is the plug region shown in light gray. The $\alpha-$lines from $AB$ to $CD$ can be found easily as discussed above; the corresponding $\beta-$lines are curved lines which are the involutes unwrapped from an imaginary co-centre circle---the evolute---with radius $\eta=\cos \left(\frac{1}{2} \cos^{-1} \tilde{\alpha}_s \right)$ shown with a dashed white line. Please note that indeed $\alpha-$lines can be introduced as tangents of this imaginary circle as well. From $CD$ to $CE$, $\alpha-$lines are spokes of a fan centred at $C$; hence, the $\beta$-lines in this part are arcs of co-centred circles at point $C$.

\begin{figure}
\centerline{\includegraphics[width=0.9\linewidth]{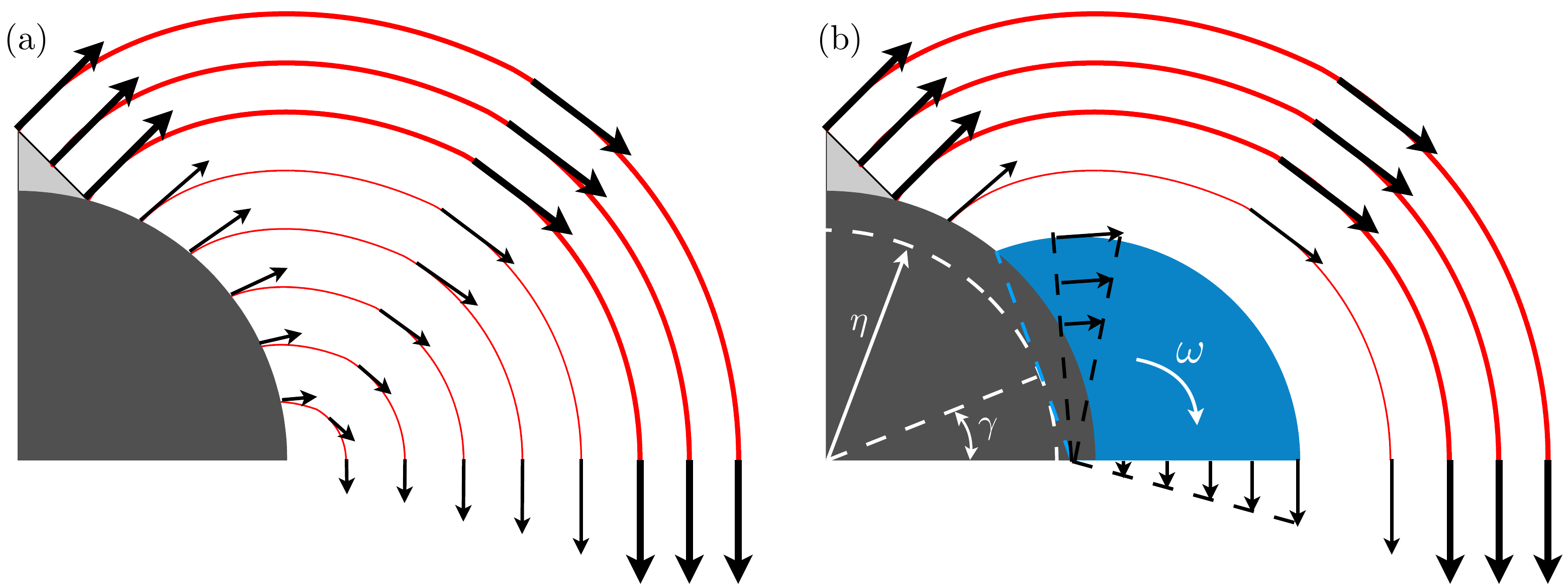}}
\caption{Schematic of the upper-bound mechanisms: (a) associated with the sliplines \citep{randolph1984limiting}, (b) combined mechanism \citep{martin2006upper} with the blue region rotating as an unyielded block. Black arrows show velocity vectors. For a better representation, in this schematic the pile velocity is assumed to be $\boldsymbol{e}_y$. Nevertheless, it should be noted that the flow has fore-aft symmetry.}
\label{fig:schematic_upperbound}
\end{figure}

Hencky's equations \citep{hill1998mathematical},
\begin{equation}
\tilde{p}+2B\theta=const. ~\text{along an $\alpha-$line}~~\&~~\tilde{p}-2B\theta=const. ~\text{along a $\beta-$line}
\end{equation}
provide the tool for finding the admissible stress field associated with these sliplines which can be used for calculating the lower bound of the plastic drag coefficient. Here, $\theta$ is the counterclockwise orientation of the $\alpha-$lines made with the $x$-axis (aligned with $OE$). The individual stress contributions can be written as,
\begin{equation}
\tilde{\sigma}_x = -\tilde{p}-B \sin (2\theta),~~\tilde{\sigma}_y = -\tilde{p}+B \sin (2\theta),~~\tilde{\tau}_{xy} = B \cos (2\theta)
\end{equation}
in the new curvilinear coordinates $\alpha$ and $\beta$. Then the lower bound can be calculated as,
\begin{equation}
\left[ C_{D,c}^p \right]_L = \frac{\left[ \tilde{F} \right]_L}{\ell_{\bot} B} = \frac{\left[ \tilde{F} \right]_L}{2B} = \int_{\partial X} \left( \tilde{\ubsigma} \boldsymbol{\cdot} \boldsymbol{n} \right) \boldsymbol{\cdot} \boldsymbol{e}_y ~\text{d}S
\end{equation}
which basically consists of four individual contributions in the shown quadrant: shear force acting on $AB$, $B \sin (\psi/2)$; shear force acting on $AC$, $B \sin \psi \cos(\psi/2)$; normal force acting on $AB$, $B \left( c + \frac{3\pi}{2} \right) \sin\left(\frac{\psi}{2}\right)$; and normal force acting on $AC$, $B \left\{ c \left[ 1 - \sin\left(\frac{\psi}{2}\right) \right] + \left[ 2 \cos\left(\frac{\psi}{2}\right) + \frac{\pi}{2} + \psi + \cos\psi - \left( \frac{3\pi}{2} + \cos\psi \right) \sin\left(\frac{\psi}{2}\right) \right] \right\}$, where $cB$ is the reference mean stress, $\left( \tilde{\sigma}_x + \tilde{\sigma}_y \right)/2$, on $CE$ (position of the centre of the Mohr's circle). Please note that $c$ will be eliminated if all four quadrants are taken into account. Hence, the total lower bound of the plastic drag coefficient will be,
\begin{equation}\label{eq:LB}
\left[ C_{D,c}^p \right]_L = \pi + 2 \psi + 2 \cos \psi + 4 \left( \cos \frac{\psi}{2} + \sin \frac{\psi}{2} \right).
\end{equation}
The upper bound calculation can be performed by postulating admissible velocity fields. Geiringer equations \citep{hill1998mathematical} make it feasible via slipline solution,
\begin{equation}
\text{d}\tilde{u}_{\alpha} - \tilde{u}_{\beta} \text{d}\theta=0 ~\text{along an $\alpha-$line}~~\&~~\text{d}\tilde{u}_{\beta} + \tilde{u}_{\alpha} \text{d}\theta=0 ~\text{along a $\beta-$line},
\end{equation}
where $\tilde{u}_{\alpha}$ and $\tilde{u}_{\beta}$ are the velocities in the directions $\alpha$ and $\beta$ in the new curvilinear coordinate system. Based on these quations, \cite{randolph1984limiting} proposed a mechanism (see figure \ref{fig:schematic_upperbound}(a)) in which velocity along $\alpha-$lines is zero and along each $\beta-$line is a constant which can be found by considering no-penetration condition on the pile or fore-aft plugs surface. As mentioned above, this mechanism leads to a relatively large uncertainty with the lower bound solution for $\tilde{\alpha}_s \neq 1$. \cite{martin2006upper} improved this mechanism by substituting a rotating rigid block (blue region in figure \ref{fig:schematic_upperbound}(b)) in the middle of the domain with angular velocity $\omega$ compatible with the no-penetration boundary condition on the pile surface; the centre of that is determined by the angle $\gamma$ and the imaginary evolute circle of radius $\eta$. By optimizing for $\gamma$, one can markedly reduce the uncertainty between the lower and upper bounds of $C_{D,c}^P$.

It is worth mentioning that although $\tilde{\alpha}_s$ and $\alpha_s$ do not have exactly the same physical origins, in what follows, we show that both display the same effect in the current problem and can be used interchangeably. Nevertheless, we should note that, $\alpha_s B$ in the viscoplastic fluid mechanics context not only marks the boundary between the presence of sliding or no-slip flow on the solid surface, but actually also controls the yield limit (see figure \ref{fig:Schematic_slip}): for instance in the case of $\alpha_s =1$, the limit of flow/no-flow reduces to having stress as large as $B$ on the solid surfaces. For sure, when there is a flow, it slides over solid boundaries depending on $\beta_s$ in this case. Yet, the case of $\alpha_s=1$ shares the same yield limit characteristics with the no-slip condition. In the perfectly-plastic mechanics context, $\tilde{\alpha}_s B$ roughly has the same interpretation since if the tangential shear stress on the solid boundary is less than that, then there is no deformed regions. Indeed, this is the strict condition at the yield limit in a deformed perfectly-plastic solid.

\subsection{Results}

\begin{figure}
\centerline{\includegraphics[width=\linewidth]{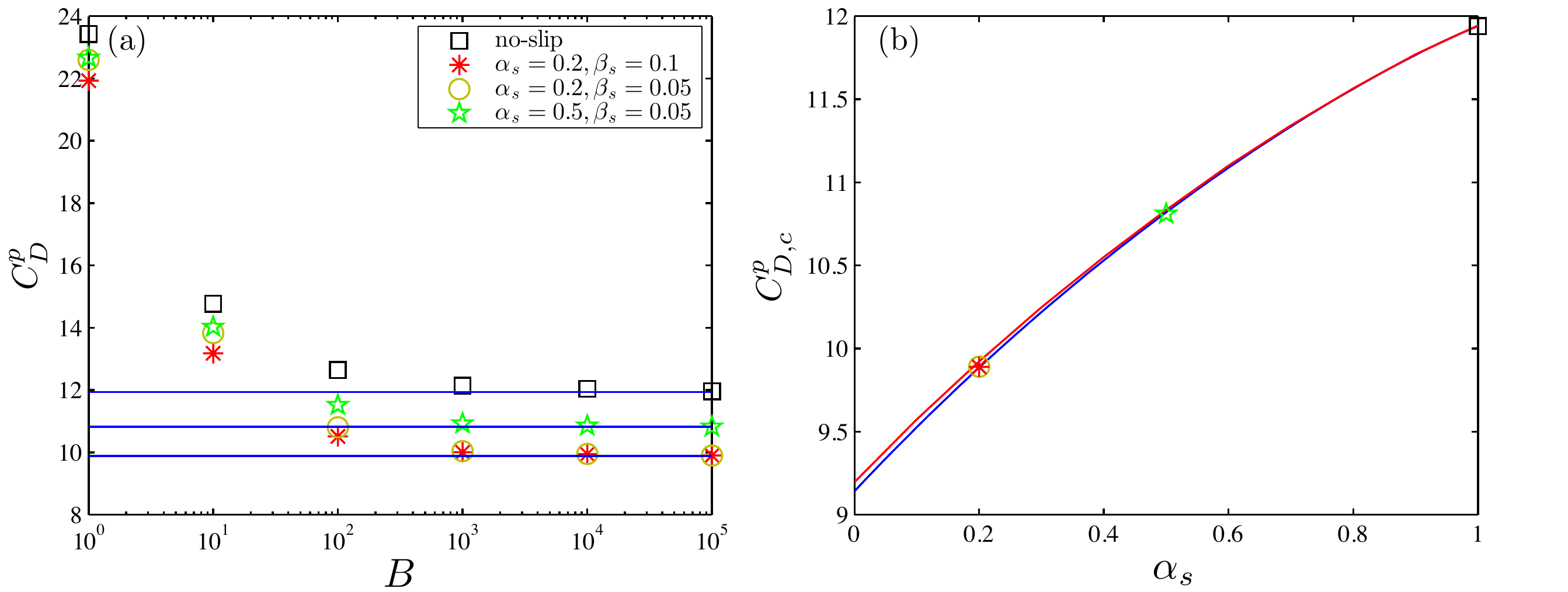}}
\caption{(a) Plastic drag coefficient versus the Bingham number for different $\alpha_s$ and $\beta_s$. The continuous blue lines are lower-bound estimations for $\tilde{\alpha}_s =$ 0.2, 0.5, and 1. (b) Critical plastic drag coefficient versus $\alpha_s$. The blue line is the lower-bound estimation and the red line shows the upper-bound estimation by the combined mechanism \citep{martin2006upper} with the same symbol interpretation as panel (a).}
\label{fig:CD}
\end{figure}

Figure \ref{fig:CD}(a) represents the plastic drag coefficient versus the Bingham number under different conditions. As shown by \cite{putz2010creeping} and \cite{chaparian2017yield}, with the no-slip condition on a circular particle surface, $\lim_{B \to \infty}C_D^p = C_{D,c}^p \approx 11.94$ which is the value that can be obtained from the expression (\ref{eq:LB}) as well when $\tilde{\alpha}_s=1$. Three main points can be drawn from figure \ref{fig:CD}(a) as follows:
\begin{itemize}
\item[(i)] Plastic drag coefficients associated with the sliding flows over the particle surface are smaller compared to the same Bingham number flows with the no-slip condition. This seems intuitive and as mentioned has been validated experimentally by \cite{jossic2001}.
\item[(ii)] The plastic drag coefficient decreases by increasing Bingham number and reaches a plateau when $B \to \infty$. More interestingly, the critical plastic drag coefficient associated with the flows of the same $\alpha_s$ asymptote to a same limiting value (see asterisks and circles). Indeed, $C_{D,c}^p$ is only a function of $\alpha_s$. However, in the Newtonian limit, $B \to 0$, this is $\beta_s$ which controls $C_D^p$ (see stars and circles).
\item[(iii)] The lower-bound estimations from the slipline solutions accurately predicts $C_{D,c}^p$. 
\end{itemize}

\begin{figure}
\centerline{\includegraphics[width=0.9\linewidth]{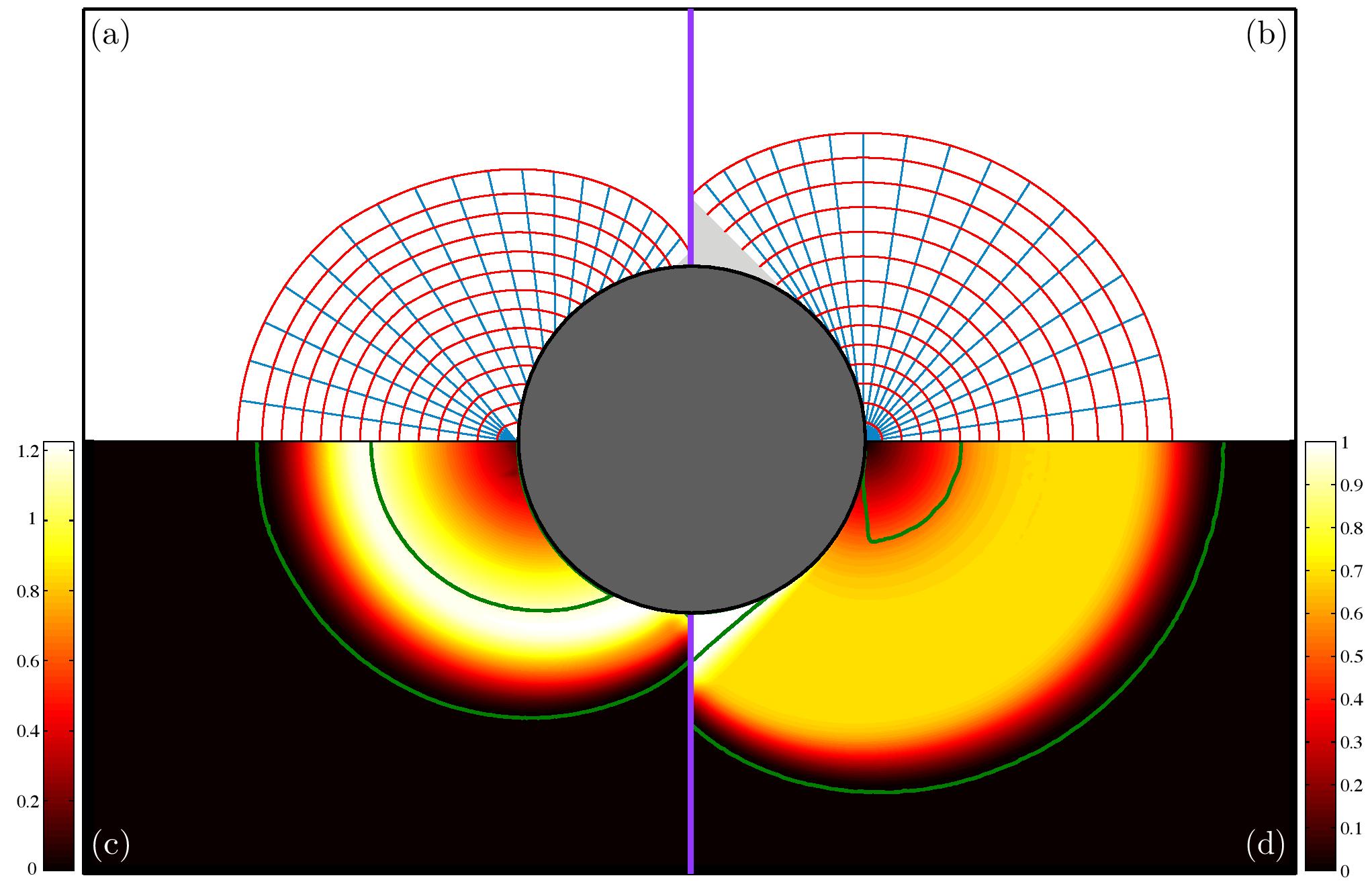}}
\caption{Sliplines for the case (a) $\tilde{\alpha}_s=0.2$ and (b) no-slip ($\tilde{\alpha}_s=1$); $\alpha-$lines in blue and $\beta-$lines in red while the rigid caps stock to the particle surface represented in light gray. The other panels show the contour of $\vert \boldsymbol{u}^* \vert$ at $B=10^5$: (c) $\alpha_s=0.2~\&~\beta_s=0.1$ and (d) no-slip flow. Green lines in panels (c,d) represent the yield usrfaces.}
\label{fig:ContourSlip}
\end{figure}

Figure \ref{fig:CD}(b) compares the lower- and upper-bounds of the critical plastic drag coefficient with the data obtained by the numerical computations. It clearly shows that even within the sliding flow context, perfectly-plastic theories could be useful in investigating the yield limit. It is worth mentioning that as investigated by \cite{chaparian2017yield} in detail, although the computations and slipline analysis display same behaviours to some extent for the no-slip case, the envelope of the characteristics does not agree fully with the yield surfaces in the viscoplastic computations. Both characteristics solution and the viscoplastic computation predict rigid caps fore and aft of the particle (although they are not exactly the same size), whereas the viscoplastic solution also has {\it kidney} plug regions along the side of the cylinder. The velocity fields are not exactly equal either; see figure 4 of the mentioned reference. The perfectly-plastic velocity has discontinuities which are diffused in the viscoplastic field by the narrow viscous boundary layers. Given the discrepancies in both stress and velocity fields for the no-slip case, it is perhaps interesting that slipline solutions still predicts the yield limit very well. Figure \ref{fig:ContourSlip} reveals that these differences exist in the sliding flows as well. Panel (a) sketches the sliplines for the case $\tilde{\alpha}_s=0.2$ and panel (c) shows the computed contours of velocity for the case $\alpha_s=0.2~\&~\beta_s=0.1$ at $B=10^5$. Panels (b) and (d) make the comparison possible with the no-slip condition. The frontal caps are significantly smaller for $\alpha_s=0.2$ while the side {\it kidneys} are bigger and are in touch with the particle surface.

\begin{figure}
\centerline{\includegraphics[width=\linewidth]{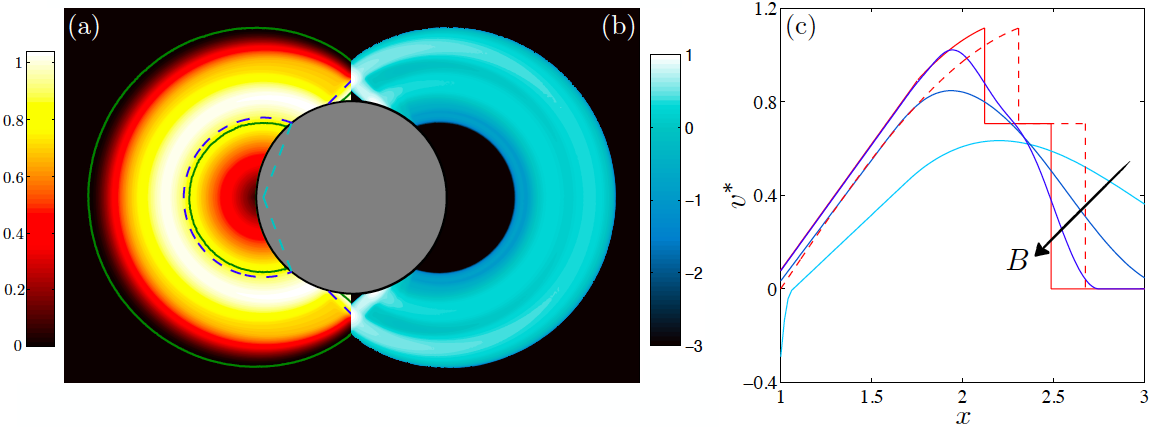}}
\caption{Features of the flow about a particle $\alpha_s=0.5~\&~\beta_s=0.05$: (a) velocity contour ($\vert \boldsymbol{u}^* \vert$) at $B=10^5$; the green lines show the yield surfaces and blue discontinuous lines are yield surfaces predicted in combined upper-bound mechanism \citep{martin2006upper} with $\eta=0.866~\&~\gamma=20.8 \degree$ and cyan discontinuous lines mark the start and end spokes of the rigid zones. (b) contour of $\log_{10} (\Vert \dot{\ubgamma}^* \Vert)$. (c) velocity profile comparison in simulations with increasing Bingham number ($B=10, 100, \text{and}~ 10^5$ from cyan to dark blue color, respectively) and upper-bound mechanisms (\cite{martin2006upper} combined mechanism with continuous red line and \cite{randolph1984limiting} mechanism with discontinuous dashed red line) over the horizontal symmetry line. Please note that particle surface velocity is $v^*=-1$.}
\label{fig:alpha05}
\end{figure}

Figure \ref{fig:alpha05} closely monitors these differences for the case $\alpha_s=0.5$. Panel (a) shows the contour of velocity in the viscoplastic problem with yield surfaces in continuous green, whereas `yield surfaces' predicted by the combined upper-bound mechanism \citep{martin2006upper} are shown with discontinuous blue lines. Moreover, panel (c) compares velocity distributions along the horizontal symmetry line: the two red lines represent the ones associated with the upper-bound mechanisms by \cite{martin2006upper} (continuous) and \cite{randolph1984limiting} (discontinuous), while the computed velocities for different Bingham numbers are shown in blue colours; the darker the Bingham number higher. As $B$ increases, the velocity distributions get closer to the ones predicted by the upper-bound mechanisms, especially the combined mechanism where even the slip velocity on the particle surface at the equator is predicted correctly. Nevertheless, again the discontinuities in the perfectly-plastic solution are replaced by the viscous boundary layers in the viscoplastic solutions.

\section{Sliding flow in model porous media}\label{sec:porous}

Non-Newtonian fluid flows through porous media, especially yield-stress fluids, are of great practical importance for numerous industries such as filtration. Due to the complexities of these type of geometries and their randomness, it is expensive to conduct numerical simulations/experiments in the full scale. Hence, a common way is to model the medium as arrays of cylinders or other model geometries. For instance, \cite{chevalier2013darcy} studied the flow of a Herschel-Bulkley fluid through confined
packings of glass beads experimentally and proposed a semi-empirical
model for the pressure drop versus the flow rate. \cite{chaparian2019porous} and \cite{de2018elastoviscoplastic}, recently, studied the effect of elasticity of the yield-stress fluid in flows through model porous media. Another related topic is the flow along uneven channels: in a series of studies Roustaei {\it et al.}~ \citep{roustaei2013occurrence,roustaei2015residuala,roustaei2015residualb,roustaei2016non} investigated in details the yield-stress fluid flow inside fractures and washouts for a wide range of flow configurations. They showed that large-amplitude variations in the duct walls lead to the formation of static unyielded zones (termed as ‘fouling layers’) adjacent to the walls. Lubrication approximation (which is based on a constant pressure gradient along the channel length), hence, is an inadequate approach to study yield-stress fluid flows along wavy channels.

It should be mentioned that all the above studies are limited to the no-slip condition. Here, in the continuation of the previous sections, we investigate the effect of slip on the pressure-driven yield-stress fluid flows in porous media.

\begin{figure}
\centerline{\includegraphics[width=0.6\linewidth]{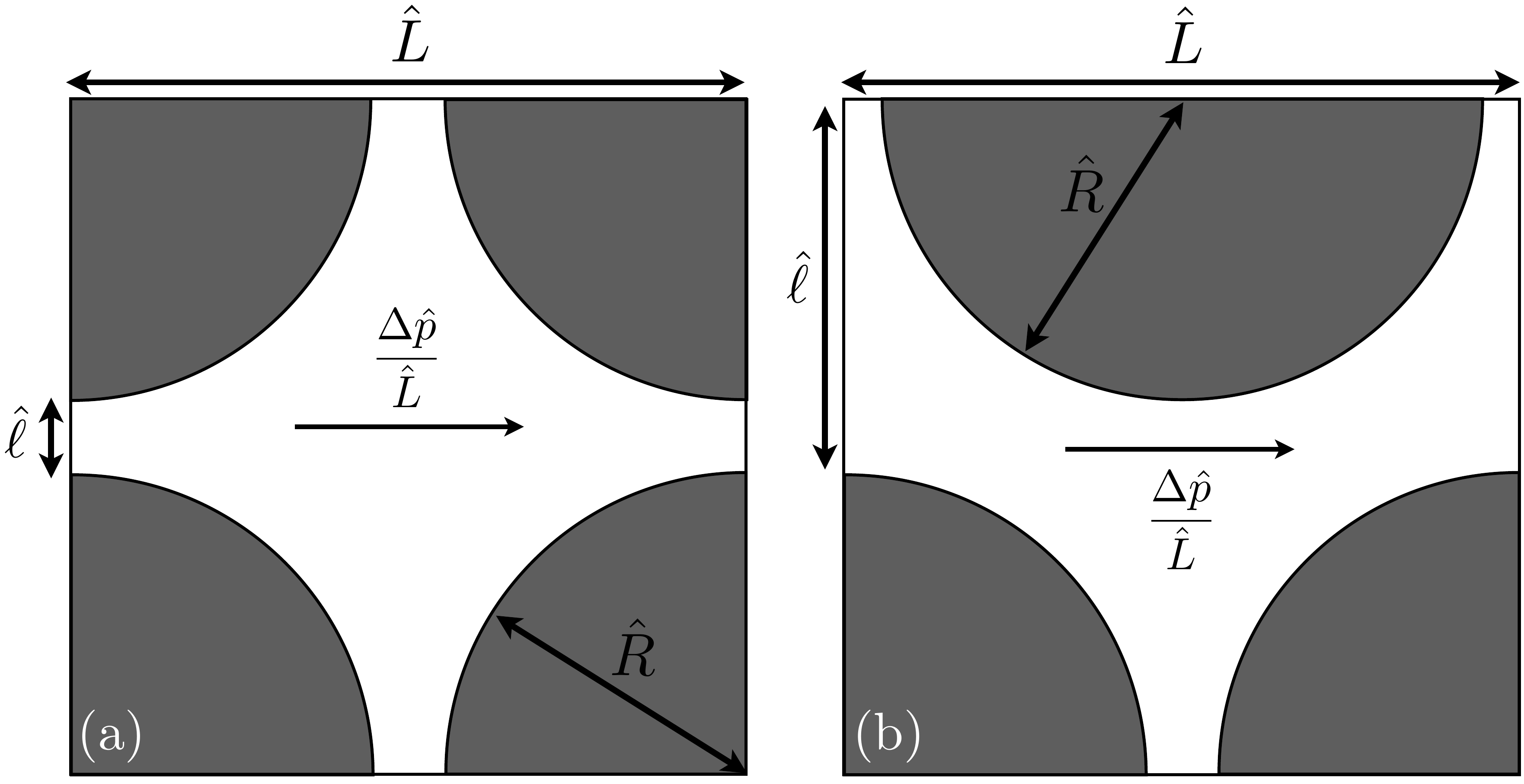}}
\caption{Schematic of the model porous media: (a) regular geometry, (b) staggered geometry.}
\label{fig:SchematicPorous}
\end{figure}

In this section, we consider flows through two model geometries mimicking porous media; the same ones as studied by \cite{chaparian2019porous}. The schematic is represented in figure \ref{fig:SchematicPorous}: both geometries are designed to have the same porosity $\phi=0.38$ (i.e. $\hat{\ell}/\hat{R} = 0.25$ and 1.25 in panels (a) and (b), respectively). The obstacles radius, $\hat{R}$, is used as the length scale. We use the [R] formulation in this section; hence, for all the cases of the same geometry, the flow rate is constant and the pressure drop is a function of the Bingham number $\left( \frac{\hat{\tau}_y \hat{R}}{\hat{\mu} \hat{U}} \right)$, $\alpha_s$, and $\beta_s$. We consider an inertialess fully developed flow with the sliding boundary condition (\ref{SlipLaw-R}) on the solid topologies ($\partial X$). Fixed flow rate condition is enforced using a Lagrange multiplier and periodic boundary conditions are applied at the inlet and outlet. On the top and bottom sides of the computational cell (horizontal faces), a symmetry boundary condition is imposed. For more details about the computational method and its validations please see \cite{chaparian2019porous}. 

\begin{figure}
\centerline{\includegraphics[width=\linewidth]{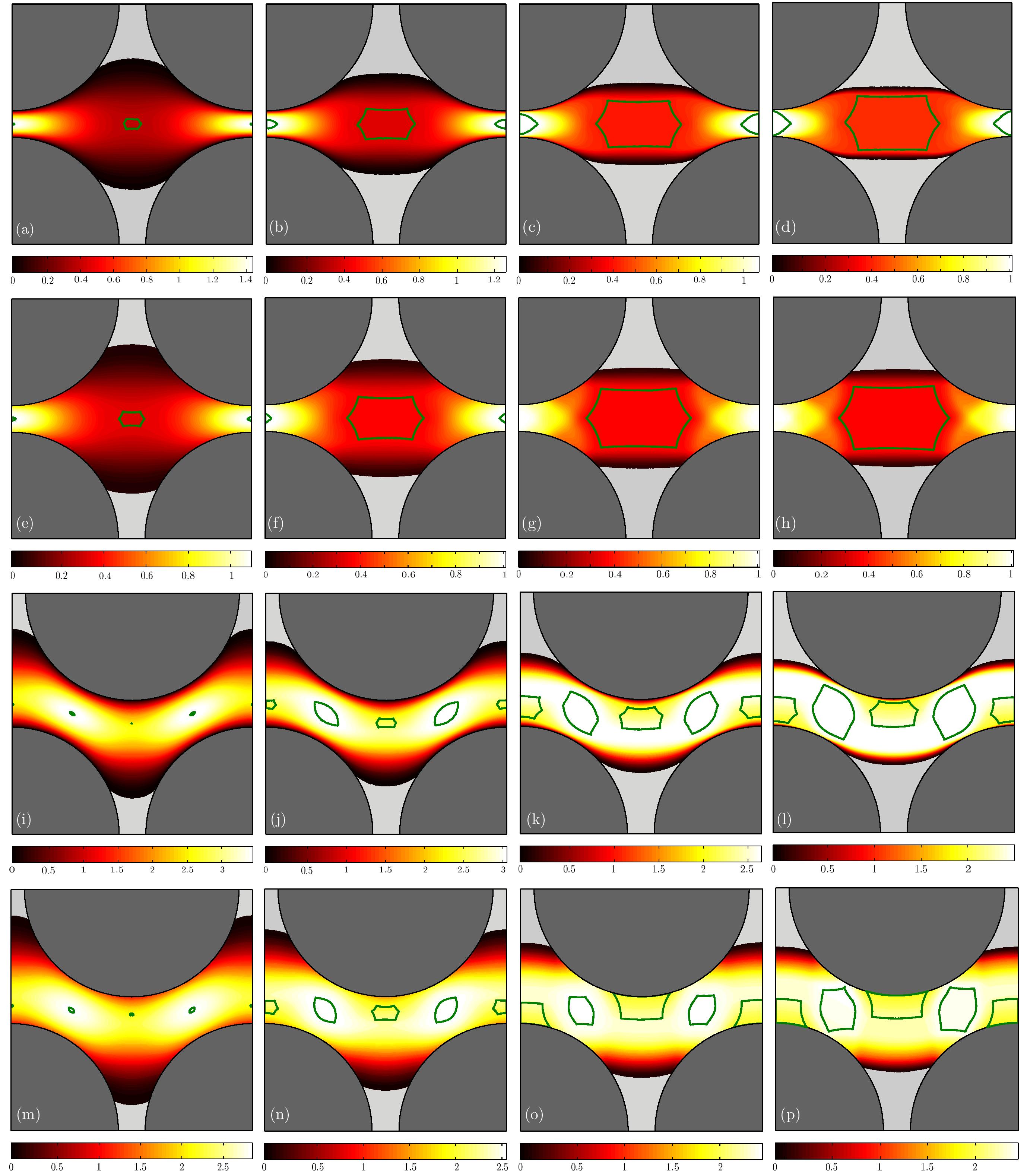}}
\caption{Velocity magnitude contour $\vert \boldsymbol{u}^* \vert$: Panels (a-d) and (i-l) show the no-slip flow in the regular and staggered geometries, respectively. Panels (e-h) and (m-p) illustrate the sliding flow ($\alpha_s=0.2,~\beta_s=0.1$) in regular and staggered geometries, respectively. Columns from left to right are associated with $B=1, 10, 100,$ and 1000, respectively. The dead zones are shown in light grey and green lines represent the yield surfaces in the middle of the flow passage.}
\label{fig:Normal_Stagg}
\end{figure}

Figure \ref{fig:Normal_Stagg} compares different flows within a wide range of Bingham numbers. Dead zones (shaded in light grey) are smaller in the sliding flows since the fluid slides over the solid surfaces and ``erodes" the fouling parts. However, the unyielded regions in the middle of the passage (yield surfaces in green) are mostly larger compared to the no-slip condition, which is intuitive. Interestingly, because of the slip, unyielded zones are prone to attach to the solid surfaces (\eg see figure \ref{fig:Normal_Stagg}(p)).

\begin{figure}
\centerline{\includegraphics[width=\linewidth]{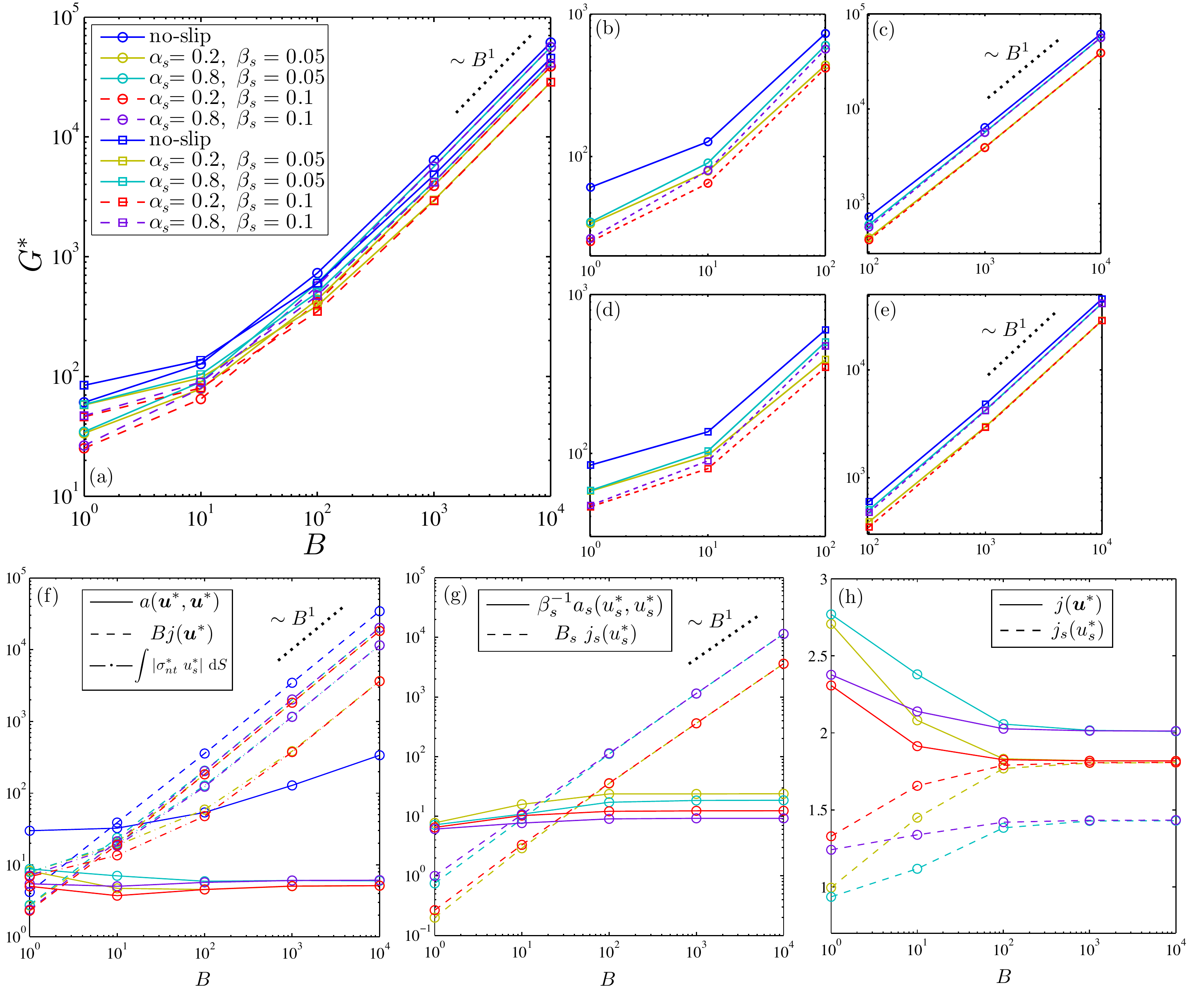}}
\caption{Features of the flow in model porous media: ~(a) $G^*$ versus $B$; the circles stand for the regular geometry and squares for the staggered cases, ~(b,c) zoom in of the panel (a) at small and large Bingham numbers, respectively (only data of the regular geometry is presented), ~(d,e) zoom in of the panel (a) at small and large Bingham numbers respectively (only data of the staggered geometry is presented), ~(f) different contributions to the total dissipation versus the Bingham number, ~(g) different contributions to the slip dissipation versus the Bingham number, ~(h) Normalized plastic dissipation and `plastic' slip dissipation versus $B$. The scale of the vertical axis in this panel is not logarithmic; it is linear. Please note that in panels (f-h) only data of the regular geometry is presented and colours interpretation is the same as the panel (a).}
\label{fig:dpdxMeccanica}
\end{figure}

Figure \ref{fig:dpdxMeccanica} reveals various features of the flow in the model porous media quantitatively. Panel (a) sketches $G^*$ with respect to the Bingham number; zoomed-in insets are provided in panels (b-e). In the range of small Bingham numbers---Newtonian limit---the pressure gradient is a function of $B$ and $\beta_s$ alone; see panels (b) and (d) in which the cyan and yellow curves ($\beta_s = 0.05$) converge together and also red and purple curves ($\beta_s = 0.1$) together. On the other hand, in the limit of large Bingham numbers (i.e. no flow or yield limit; panels (c) and (e)), the critical pressure gradient is a function of $B$ and $B_s$ or indeed $B$ and $\alpha_s$: yellow and red curves with $\alpha_s=0.2$ are almost indistinguishable; the same for the purple and cyan curves with $\alpha_s=0.8$. Moreover, as shown by \cite{chaparian2019porous}, in the yield limit ($B \to \infty$), the pressure gradient scales with $\sim B$; this is true for both the no-slip and sliding flows, yet with different pre-factors. Indeed, as shown in \S \ref{sec:R} for the simple channel flow,
\[
\lim_{B \to \infty} \frac{G^*(\text{no-slip})}{G^*(\alpha_s)}=\frac{1}{\alpha_s},
\]
which is the case in the complex flows through model porous media as well.

Panel (f) confirms that the viscous dissipation, even in sliding flows, is at least one order of magnitude less than the plastic and slip dissipations. Interestingly, for sliding flows, $a(\boldsymbol{u}^*,\boldsymbol{u}^*)$ converges to its limiting value when $B \to \infty$ and the limiting value is a function of $\alpha_s$ only. Whereas, in the no-slip flow, the viscous dissipation increases by increasing Bingham number but with a much smaller rate compared to the plastic dissipation. Panel (f) also shows that the leading order dissipation terms (i.e. $Bj(\boldsymbol{u}^*)$ and $\int \vert \sigma^*_{nt} u^*_s \vert~\text{d}S$) scales with $\sim B$ at the yield limit for sliding flows which is the main reason of yielding to the same scaling for $G^*$ at this limit; see expression (\ref{energyR}). Panel (g) also establishes that the leading order term in the slip dissipation is the `plastic' slip dissipation $B_s j_s(u^*_s)$.

Please note that although $j(\boldsymbol{u}^*)$ of the no-slip flow is not shown in panel (h), the trend is the same as the sliding flows, but of course with larger values as it is clear from panel (f) where $Bj(\boldsymbol{u}^*)$ is shown. 

Under the no-slip condition, the asymptotic values of $j(\boldsymbol{u}^*)$ at $B \to \infty$ are equal to $3.42$ and $12.57$ for regular and staggered geometries which yield to $Od_c = \frac{\hat{\tau}_y}{\hat{R} \hat{G}_c} \approx 0.1645$ and $\approx 0.2237$, respectively, using expression (\ref{OdcBtoInf}) in the absence of any slip dissipation. The critical Oldroyd number in the presence of slip is calculated using the same expression in Table \ref{tab:OdcModel}. Please note that as an approximation, one may alternatively use,
\begin{equation}\label{eq:jsapprox}
Od_c = \lim_{B \to \infty} \frac{\displaystyle \int_{\Omega \setminus \bar{X}} ~ \boldsymbol{u}^* \boldsymbol{\cdot} \boldsymbol{e}_x ~\text{d}A}{j(\boldsymbol{u}^*)+\displaystyle \frac{1}{B} \int_{\partial X} \vert \sigma^*_{nt} ~u^*_s \vert ~\text{d}S} \approx \displaystyle \lim_{B \to \infty} \frac{\ell L}{(1+\alpha_s)~j(\boldsymbol{u}^*)},
\end{equation}
in which $j_s(u_s^*)$ is approximated by $j(\boldsymbol{u}^*)$ as is suggestive by panel (h), especially for small values of $\alpha_s$ (i.e. more slippery flows). These values are reported in Table \ref{tab:OdcModel} as well. Negligible difference between $Od_c$ calculated using expressions (\ref{OdcBtoInf}) and (\ref{eq:jsapprox}) for $\alpha_s=0.2$ shows the validity of this approximation when the `sliding yield stress' is much smaller than the material yield stress.

\begin{table}
  \begin{center}
\def~{\hphantom{0}}
  \begin{tabular}{lcccc}
        & Reg. (exact)  &  Reg. (exp. (\ref{eq:jsapprox})) & Stagg. (exact)  &  Stagg. (exp. (\ref{eq:jsapprox})) \\[3pt]
       no-slip & 0.1645 & --- & 0.2237 & --- \\
       $\alpha_s=0.8$   & 0.1781 & 0.1553 & 0.2427 & 0.2125\\
       $\alpha_s=0.2$   & 0.2581 & 0.2578 & 0.3502 & 0.3486\\
  \end{tabular}
  \caption{Values of $Od_c$ for regular and staggered geometries}
  \label{tab:OdcModel}
  \end{center}
\end{table}

We shall mention that as it is clear from figure \ref{fig:dpdxMeccanica}(c,e,h), $Od_c$ only depends on $\alpha_s$ as reported in Table \ref{tab:OdcModel} (at least for $0 \leqslant \beta_s \leqslant 0.1$).

\section{Towards more complex geometries}

Although previous studies on the yield-stress fluid flows in the model porous media uncovered some generic interesting aspects of the problem, yet they were not capable of capturing the whole physics behind these type of complex flows. For instance, \cite{talon2013determination} utilized the Lattice Boltzmann Method (LBM) to study the Bingham fluid flow in an stochastically reconstructed porous media. These authors showed that due to the yield stress, channelization can happen: the material will flow only in self-selected paths depending on the imposed driving pressure gradient; at the yield limit it is only one path and by increasing the pressure gradient gradually, other paths will open up. This has been supported by the work of \cite{liu2019darcy} in model pore networks connected by straight tubes.

To capture this kind of fascinating and more realistic behaviours, in this section, we will complement the findings in \S \ref{sec:porous} by studying the flow in random-designed porous media with a wide range of porosities. The computational domain is a box of dimensions $50 \times 50$ where the fluid flows around 2D fixed rigid circles of radius unity with symmetric boundary conditions on the pair of the horizontal and vertical box faces. Again we use the [R] formulation in this section and the same numerical method as in the previous section. To have a better comparison, in addition to the random porous media, we simulate the flow again in the two introduced model porous media (figure \ref{fig:SchematicPorous}); yet with the same porosity as the random cases: the size of the model cell in figure \ref{fig:SchematicPorous} is $L=\sqrt{\frac{\pi}{1-\phi}}$.

\subsection{Baseline no-slip flow}
\begin{figure}
\centerline{\includegraphics[width=\linewidth]{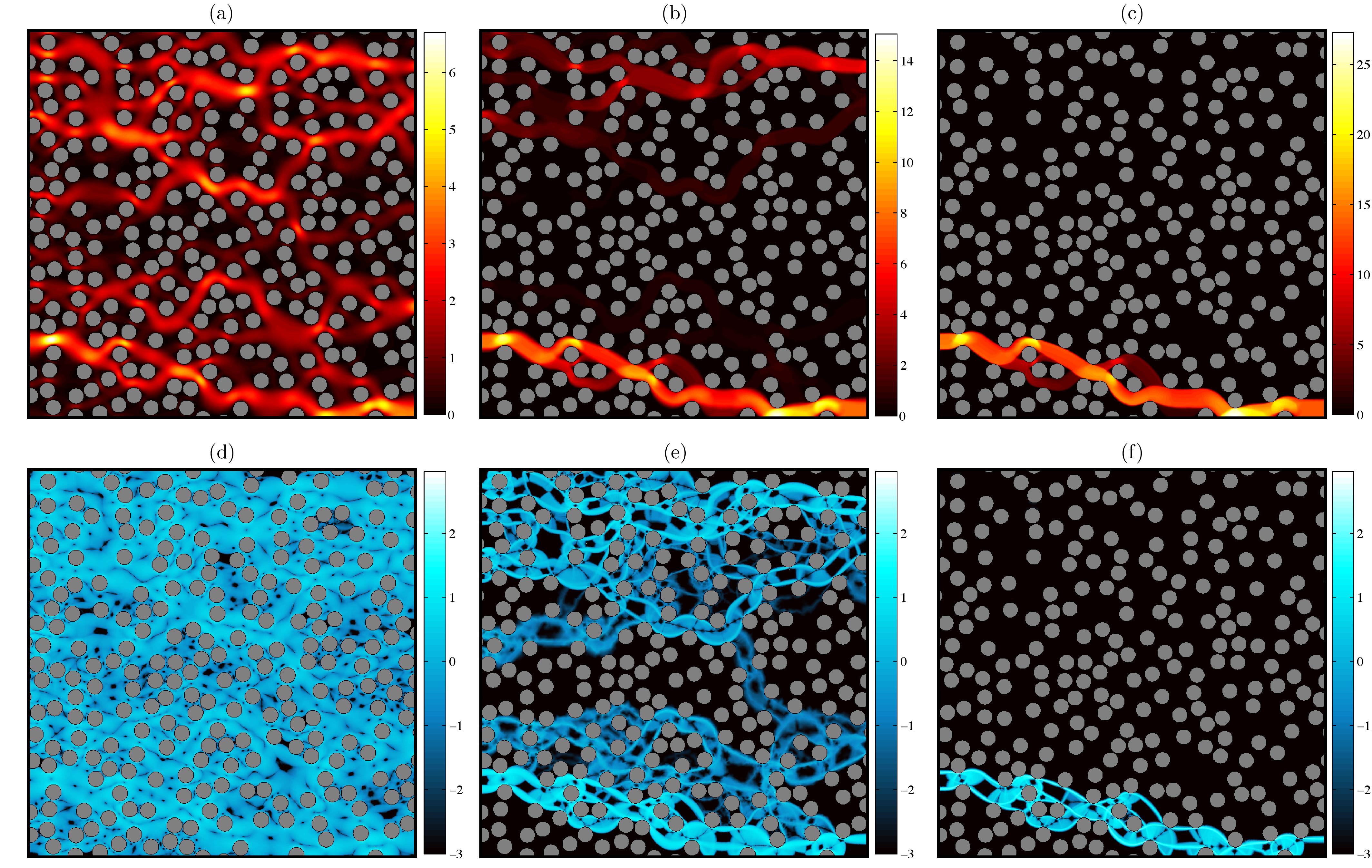}}
\caption{$\phi=0.7$: panels (a-c) illustrate $\vert \boldsymbol{u}^* \vert$ and (d-f) contour of $\log_{10} (\Vert \dot{\ubgamma}^* \Vert)$ for flows with no-slip condition. Please note that the range of colourbars are different in velocity contours whereas it is the same in the bottom row. Columns from left to right are associated with $B=1, 100$ and $10^4$, respectively.}
\label{fig:07noslip}
\end{figure}

Figure \ref{fig:07noslip} illustrates the no-slip flow in the case $\phi = 0.7$ for a wide range of Bingham numbers from unity to $10^4$. Top panels show the velocity magnitude contours and the bottom row the log-scale of the second invariant of the rate of strain tensor. When the Bingham number increases, the flow becomes more and more localized, and at the yield limit ($B \to \infty$), the flow only passes through a single channel. Please note that the black level contours represent the quiescent parts. It is worth mentioning that due to the [R] scaling, the flow rate in all the simulations is the same (for a fixed geometry regardless of the Bingham number, $\alpha_s$ and $\beta_s$), hence, in panel (c) the whole fluid passes through the single open channel which results in extremely high velocities (compare the level of contours in the top panels). By looking at the panels from left to right sequentially, the channelization characteristic is well observable: by increasing the Bingham number (i.e. moving to the yield limit), the flow
occurs in lesser pathes, until only one path remains at the yield limit.

\begin{figure}
\centerline{\includegraphics[width=0.6\linewidth]{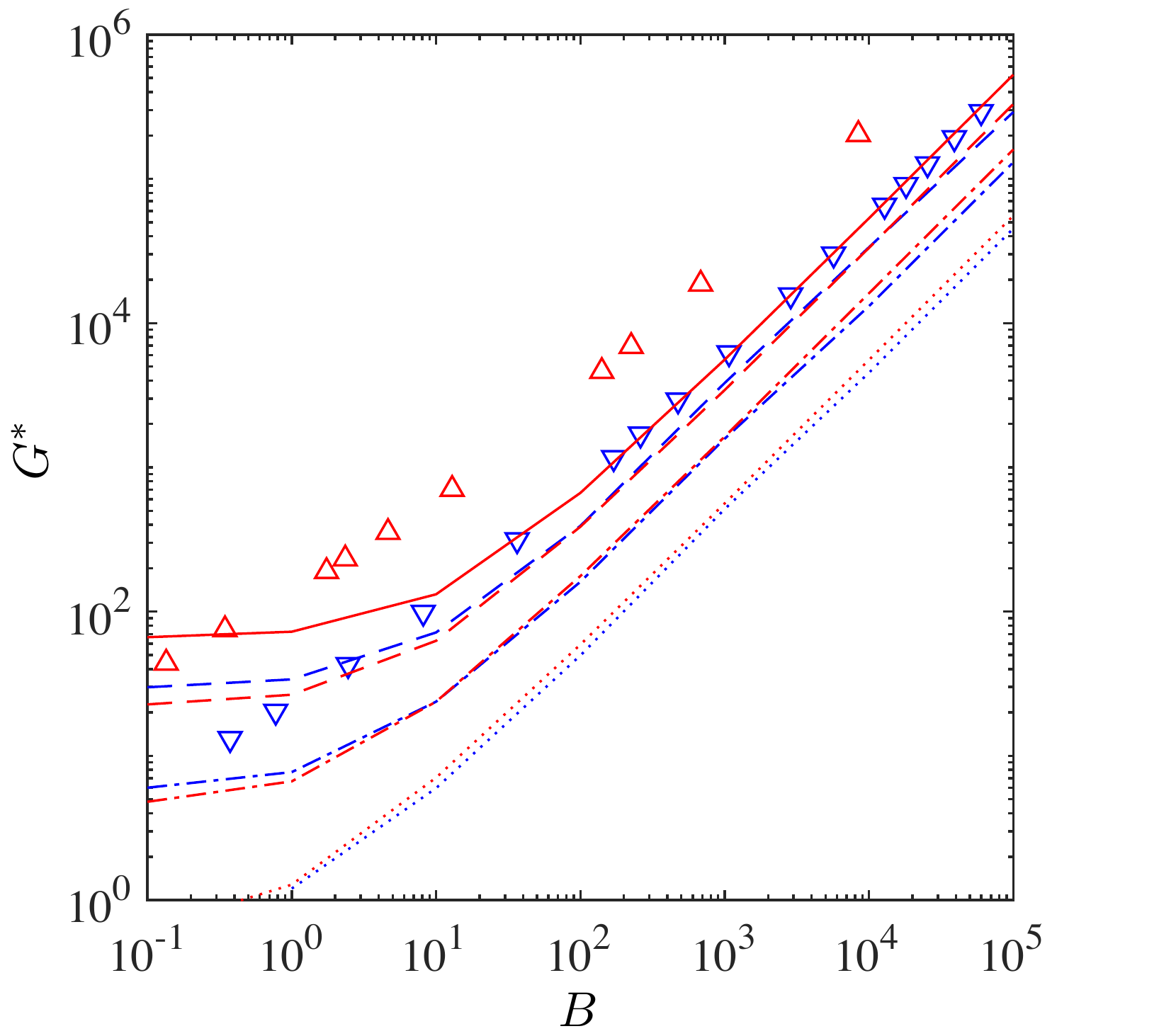}}
\caption{Pressure drop with respect to the Bingham number under the no-slip condition. The red colour stands for simulations of the model porous media and the blue colour the simulations of the random porous media. The lines represent our computations: the continuous line corresponds to $\phi=0.38$, dashed line to $\phi=0.5$, dashed-dotted line to $\phi=0.7$, and the dotted line to $\phi=0.9$. The symbols are data borrowed from \cite{bauer2019experimental}: simulations for FCC packing of spheres ($\vartriangle$; $\phi=0.218$), simulations for random array of overlapping spheres ($\triangledown$; $\phi=0.429$).}
\label{fig:dpdxnoslip}
\end{figure}

Figure \ref{fig:dpdxnoslip} shows the pressure drop against the Bingham number. The lines are computed data in the present study and the symbols are computational and experimental data by \cite{bauer2019experimental} for 3D flows. The blue lines are acquired from the randomized porous media and the red lines from the two model porous media for a wide range of porosities. To have a better comparison, we combined $G^*$ from the regular and staggered models in the red lines by averaging. This figure demonstrates that the model porous media closely predicts the pressure drop in the randomized porous media, however, there are small discrepancies, more pronounced in the two limits of the problem---Newtonian and the yield limit. Yet, in the intermediate regime of the Bingham numbers, the predictions are more satisfactory. Table \ref{tab:OdcNoslip} compares the critical Oldroyd number under the no-slip condition computed in different configurations. As can be seen in figure \ref{fig:dpdxnoslip}, the model porous media slightly over predicts the pressure drop which results in smaller $Od_c$ compared to the randomized geometries.

In figure \ref{fig:dpdxnoslip}, we also compare 2D simulations in the present study and 3D simulations by \cite{bauer2019experimental}. They are partially comparable; specifically in the large Bingham number limit where a large portion of domain is filled with static/fouling regions. However, a closer comparison in the Newtonian limit is suggestive of different trends, which could be the consequence of different flow structures in 2D and 3D flows in this limit and also the difference between the viscous functionality in the Bingham model which is used here and the Herschel-Bulkley model used by \cite{bauer2019experimental}. More sophisticated analyses/comparisons of the model and random porous media under the no-slip condition are presented in Appendix \ref{sec:flowcurve}, as the main focus of the present study is on the sliding flows.

\begin{table}
  \begin{center}
\def~{\hphantom{0}}
  \begin{tabular}{lcccc}
        & $\phi=0.38$ & $\phi=0.5$  &  $\phi=0.7$ & $\phi=0.9$  \\[3pt]
       Randomized & --- & 0.3511 & 0.7678 & 2.2096 \\
       Regular geo.   & 0.1645 & 0.2804 & 0.6163 & 1.7996 \\
       Staggered geo.   & 0.2237 & 0.3308 & 0.6365 & 1.8001 \\
  \end{tabular}
  \caption{Comparison of $Od_c$ (under no-slip condition) for different geometries}
  \label{tab:OdcNoslip}
  \end{center}
\end{table}
 
\subsection{Effect of slip}

\begin{figure}
\centerline{\includegraphics[width=\linewidth]{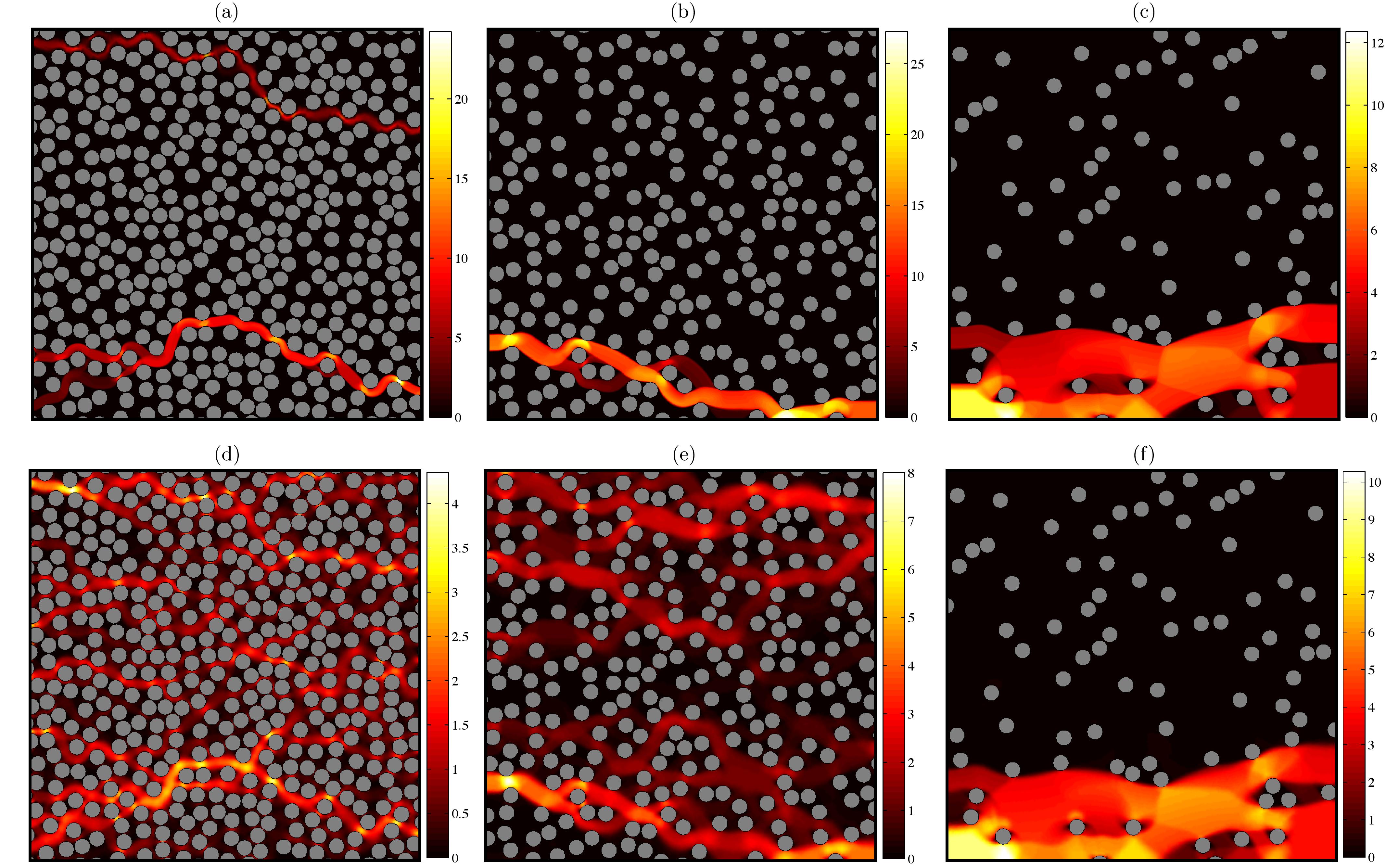}}
\caption{$\vert \boldsymbol{u}^* \vert$ at $B=10^4$: ~(a-c) no-slip flow, ~(d-f) sliding flow ($\alpha_s=0.2$ and $\beta_s=0.1$). Columns from left to right correspond to $\phi=0.5, 0.7$, and 0.9, respectively.}
\label{fig:Channelisation_u}
\end{figure}

\begin{figure}
\centerline{\includegraphics[width=\linewidth]{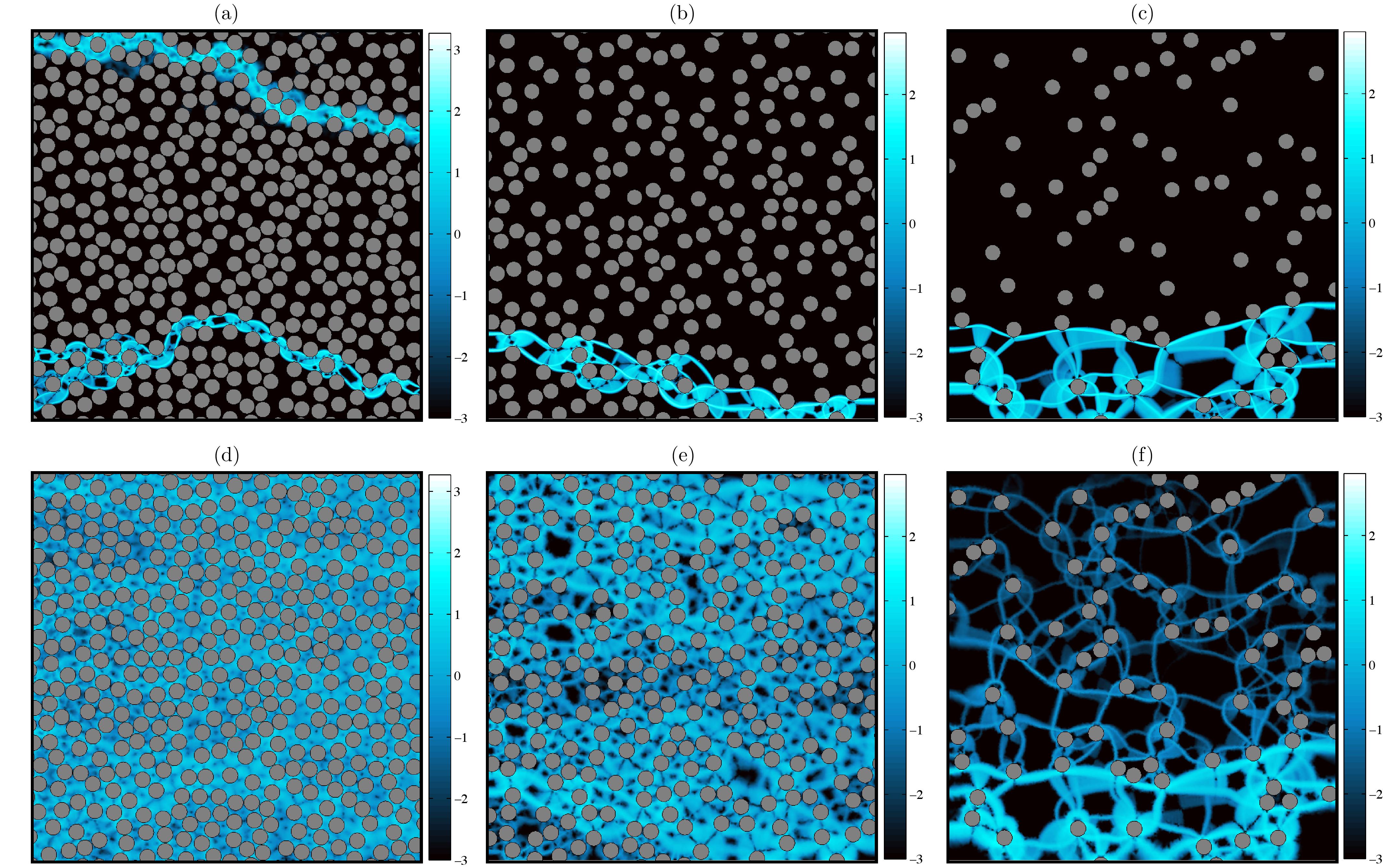}}
\caption{$\log_{10} \left( \Vert \dot{\ubgamma}^* \Vert \right)$ at $B=10^4$: ~(a-c) no-slip flow, ~(d-f) sliding flow ($\alpha_s=0.2$ and $\beta_s=0.1$). Columns from left to right correspond to $\phi=0.5, 0.7$, and 0.9, respectively. Please note that for each row, we have used the same range of colourbar.}
\label{fig:Channelisation_density}
\end{figure}

In this subsection, we take a close look at the effect of slip in the randomized porous media. Figures \ref{fig:Channelisation_u} and \ref{fig:Channelisation_density} compare the velocity and the rate of strain fields, respectively, between the flows under the no-slip condition and the sliding flows ($\alpha_s=0.2~\&~\beta_s=0.1$) at $B=10^4$. These two figures show that when the flow can slide over the pores surfaces, the channelization is not strong compared to the no-slip condition. Indeed, the flow slides over the solid surfaces which cover a large portion of the domain in the small porosities limit and consequently there is a flow in most of the passages. However, in the limit of high porosities (i.e. less solid surfaces), the effect of slip is negligible: compare panels (c) and (f) in figures \ref{fig:Channelisation_u} and \ref{fig:Channelisation_density}. The static regions under the no-slip condition and in the sliding flows are almost the same when porosity is as high as $\phi=0.9$. There are slightly sheared parts out of the main open path (see panel (f) of figure \ref{fig:Channelisation_density}), but the flow mainly passes through the open channel whose shape and location are identical to the no-slip case.

\begin{figure}
\centerline{\includegraphics[width=0.6\linewidth]{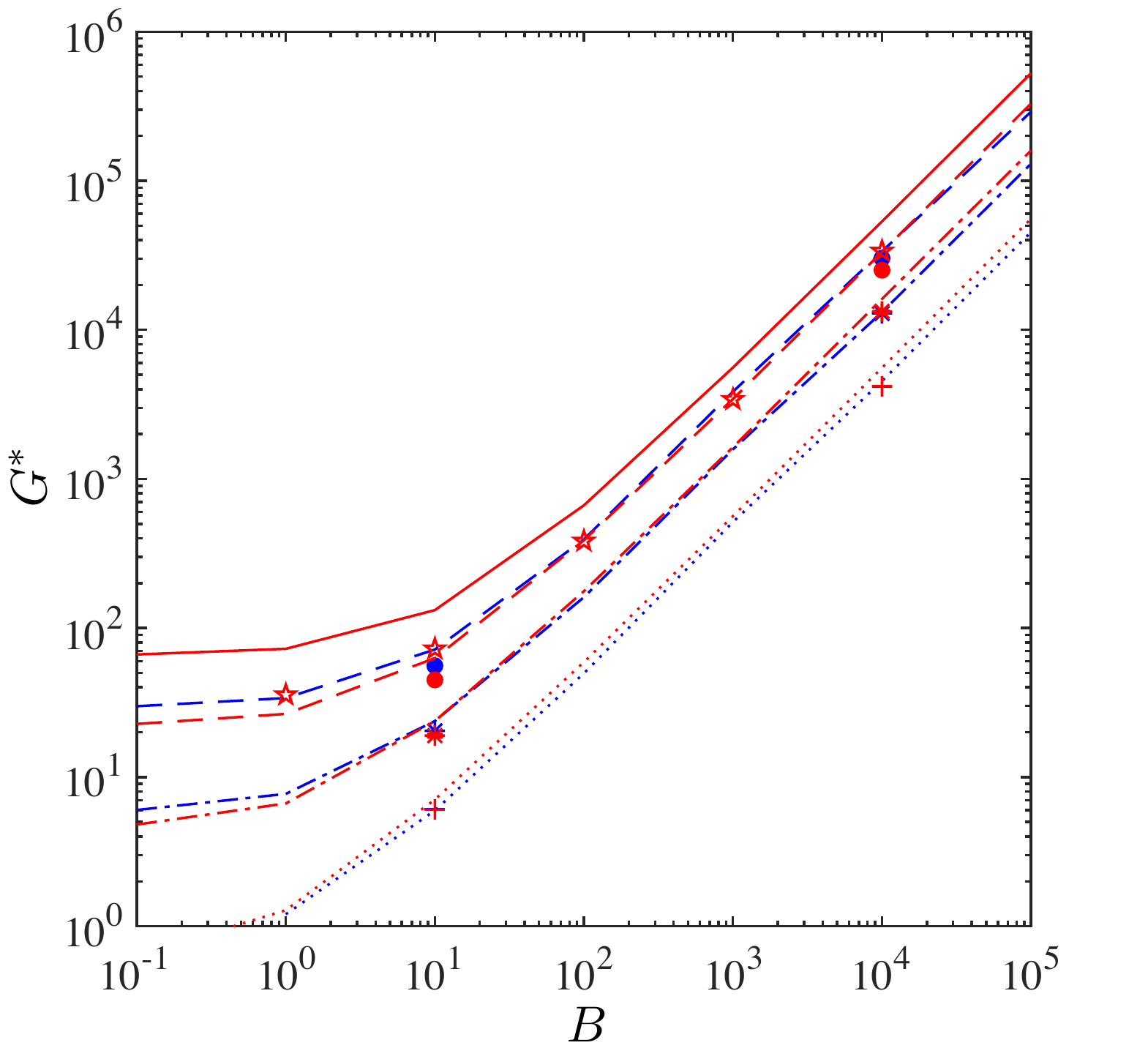}}
\caption{Pressure drop with respect to the Bingham number. The red colour stands for the simulations of the model porous media and blue colour the simulations of the random porous media. The lines are borrowed from figure \ref{fig:dpdxnoslip} (no-slip condition): the continuous line corresponds to $\phi=0.38$, dashed line to $\phi=0.5$, dashed-dotted line to $\phi=0.7$, and the dotted line to $\phi=0.9$. The symbols are computations of the sliding flow ($\alpha_s=0.2~\&~\beta_s=0.1$): stars correspond to $\phi=0.38$, full circles to $\phi=0.5$, asterisks to $\phi=0.7$, and the pluses to $\phi=0.9$.}
\label{fig:dpdxslip}
\end{figure}

Figure \ref{fig:dpdxslip} looks at this phenomenon from the pressure drop perspective. The lines (no-slip simulations) are borrowed from figure \ref{fig:dpdxnoslip} for reference. The symbols refer to sliding flows ($\alpha_s=0.2~\&~\beta_s=0.1$); the blue ones collected from the simulations in the randomized porous media and the red ones from the model porous media: please note that again we plot the average value of $G^*$ corresponding to the two model porous media (normal and staggered cases). This figure demonstrates that as the porosity increases, the effect of slip on the pressure drop decreases, in agreement with figures \ref{fig:Channelisation_u} and \ref{fig:Channelisation_density} for the model porous media. Indeed, the symbols are much closer to their corresponding lines with the same porosity as $\phi \to 1$. Moreover, in this high porosity limit, the model porous media predictions are appropriate approximations of the pressure drop in the randomized porous media: the asterisks and pluses with different colours are not distinguishable. However, in the other limit (low porosities), the effect of slip on the pressure drop is more pronounced: the pressure drop in the sliding flows are smaller compared to no-slip flows which is intuitive. Moreover, the intensity of the slip effect on the pressure drop depends on the geometric structures since there is a significant discrepancy between the predictions of the model porous media and the randomized porous media: the blue and red full circles are relatively far from each other specifically in the low Bingham number regime; please note the logarithmic scale.

\section{Summary \& discussion}\label{sec:SD}

The hydrodynamic features of sliding flows of yield-stress fluids are investigated in the present study. It has been well-documented in the literature that yield-stress fluids slide over solid surfaces due to microscopic effects: formation of a lubrication layer of solvent and elastic deformation of soft particles in the vicinity of the solid surface \citep{meeker2004PRL}. Firstly, we formulated a general vectorial form of the well-known `stick-slip' law and presented a numerical algorithm based on the augmented Lagrangian scheme combined with anisotropic mesh adaptation to attack these kind of problems. We derived some theoretical tools as well to formulate the yield limit in the presence of slip. The whole framework was benchmarked in a simple channel Poiseuille flow. Then we have moved forward to address more complicated problems including moving solid surfaces and the sliding flow over complex topologies.

Firstly, we simulated the slippery particle sedimentation problem and addressed the yield limit in detail. Slipline solutions from the perfectly-plastic mechanics were revisited and utilized to find the yield limit in the presence of slip. The slipline method has proven capable of finding the yield limit, even though the velocity/stress fields obtained from slipline solutions can differ from the viscoplastic solutions at large, yet finite Bingham numbers; see \cite{chaparian2017yield}. The conclusions from the present study are similar, showing again that the lower and upper bounds of the plastic drag coefficient are excellent estimations of the yield limit of slippery particle motion.

Secondly, we addressed the complex sliding flows in the model and randomized porous media. Different hydrodynamic features of the flow were investigated and the critical pressure gradient to ensure continuous non-zero flow was reported. Moreover, some general conclusions were drawn about modelling the flow in porous media by comparing the results computed from the model configurations and more realistic random-designed porous media; both under the no-slip condition and sliding flows. Furthermore, it was shown that the effect of slip is more severe in the low porosity limit where a larger portion of the domain is filled by occupying solids (i.e. a large portion of the flow is in contact with the solid surfaces). In this limit, the pressure drop of a slippery flow is decreased compared to the flow under the no-slip condition. For instance, the pressure drop under the no-slip condition at $\phi=0.5$ closely follows the pressure drop of the sliding flow for the case $\phi=0.38$ when $\alpha_s=0.2~\&~\beta_s=0.1$ (see figure \ref{fig:dpdxslip}). Indeed, an effective porosity can be defined for the sliding flows in porous media to predict the pressure drop: $G^* (\phi_{eff}; \text{no-slip}) = G^* (\phi; \alpha_s, \beta_s)$ where $\phi_{eff} > \phi$. In the high porosity limit, however, $\phi_{eff} \approx \phi$ since the solid surfaces are occupied a small portion of the whole domain.

By investigating different physical problems in the presence of slip from particle motion to pressure-driven flows in complex geometries, the following general conclusions can be drawn about sliding flows of viscoplastic fluids:
 
\begin{itemize}
\item[(i)] There are three source of dissipation in sliding flows: viscous, plastic and slip dissipations. Slip dissipation itself can be split into two main contributions: `viscous' slip dissipation which comes from the slip coefficient $\beta_s$ and `plastic' slip dissipation which is the contribution of the sliding yield stress.
\item[(ii)] In the yield limit, the leading order contribution to the total slip dissipation is the `plastic' slip dissipation. Exploiting that viscous dissipation is at least one order of magnitude less than the plastic dissipation, we can conclude that the yield limit of sliding flows is indeed controlled by $\alpha_s$ or sliding yield stress. This was evidenced both for particle sedimentation and pressure-driven flows in porous media.
\item[(iii)] In the other limit (Newtonian limit), however, the flow characteristics are determined by the slip coefficient $\beta_s$. For instance, the plastic drag coefficient in the particle sedimentation problem and the pressure drop in the porous media are the same for the flows with the same $\beta_s$ as $B \to 0$.
\end{itemize}

At the end we shall mention that although in the entire study we considered $k=1$ and the Bingham model as the rheological representative, but the presented framework and most of the conclusions which are drawn are independent of these assumptions. In Appendix \ref{sec:kneq1}, we will show how we can adopt the presented numerical algorithm and the analytical tools to study the cases in which $k \neq 1$ and $n \neq 1$.

\section*{Acknowledgment}
E.C. gratefully acknowledges the Linn\'e FLOW PostDoc grant during the course of this study. The authors appreciate the support of Swedish Research Council through grant VR 2017-0489. This project also has received funding from the European Research Council (ERC) under the European Union's Horizon 2020 research and innovation programme (grant agreement No. 852529).

\section*{Declaration of Interests}
The authors report no conflict of interest.

\appendix

%

\section{Analytical solution of the sliding Poiseuille flow}

We derive analytical solutions for the sliding Poiseuille flow in a channel with gap width unity considering both [M] and [R] formulations.

\subsection{{\normalfont [M]} problem}\label{app:M}
For the [M] problem, the only non-zero stress component is the shear stress which can be derived from,
\begin{equation}
\frac{\text{d}\tau_{xy}}{\text{d}y} = 1,
\end{equation}
and the slip velocity can be calculated from,
\begin{equation}
u_{{s}} = \left\{
\begin{array}{ll}
\beta_s \left( \vert \tau_w \vert - Od_s \right), & \text{iff}~~ \vert \tau_{w} \vert > Od_s, \\[2pt]
0, & \text{iff}~~ \vert \tau_{w} \vert \leqslant Od_s,
\end{array} \right.
\end{equation}
where $\tau_w$ is the wall shear stress. From the constitutive equation, we know that in the yielded regions (if there are any),
\begin{equation}
Od ~\text{sgn}(y)-y=\dot{\gamma}_{xy}=\frac{\text{d}u}{\text{d}y}.
\end{equation}
Hence, if $\vert y \vert \geqslant Od$,
\begin{equation}
u = Od~\vert y \vert-\frac{y^2}{2} + A,
\end{equation}
with the boundary condition,
\begin{equation}
u~\left(y=\pm \frac{1}{2}\right) = u_s = \beta_s \left(\frac{1}{2} - Od_s \right)
\end{equation}

Therefore, three different scenarios can happen based on values of $Od$ and $Od_s$:
\begin{equation}
u = \left\{
\begin{array}{ll}
0, & \text{iff}~~ 1/2 \leqslant Od_s ~\text{(i.e. no flow)}, \\[2pt]
\beta_s (\frac{1}{2} - Od_s), & \text{iff}~~ Od_s < 1/2 \leqslant Od ~\text{(i.e. fully sliding plug)}, \\[2pt]
u^y, & \text{iff}~~ Od < 1/2 ~\text{(i.e. deforming regime)},
\end{array} \right.
\end{equation}
where,
\begin{equation}
u^y = \left\{
\begin{array}{ll}
\frac{1}{2} \left( Od - \frac{1}{2} \right)^2 + \beta_s \left( \frac{1}{2} - Od_s \right), & \text{iff}~~ \vert y \vert \leqslant Od ~\text{(i.e. core plug region)}, \\[2pt]
Od~(\vert y \vert -\frac{1}{2}) + \frac{1}{8} - \frac{y^2}{2} +  \beta_s (\frac{1}{2} - Od_s), & \text{iff}~~ Od < \vert y \vert ~\text{(i.e. sliding sheared region)}.
\end{array} \right.
\end{equation}
Hence, $Od_c = 1 / 2 \alpha_s$.

\subsection{{\normalfont [R]} problem}\label{app:R}
In the [R] formulation, we always have a non-zero flow rate due to the velocity scaling. Hence, we shall identify two different regimes: deforming regime at small and moderate Bingham numbers ($B < \bar{B}$) and a fully sliding plug at $B \to \infty$ (or strictly when $B > \bar{B}$). The shear stress satisfies,
\begin{equation}
\frac{\text{d}\tau^*_{xy}}{\text{d}y^*} = G^*,
\end{equation}
with the slip velocity,
\begin{equation}
u_s^* = \left\{
\begin{array}{ll}
\beta_s \left( \vert \tau^*_{w} \vert - B_s \right), & \text{iff}~~ \vert \tau_w^* \vert > B_s, \\[2pt]
0, & \text{iff}~~ \vert \tau_w^* \vert \leqslant B_s.
\end{array} \right.
\end{equation}
In deforming regime, in the yielded regions,
\begin{equation}
B ~\text{sgn} (y^*)-G^* y^*=\dot{\gamma}^*_{xy}=\frac{\text{d}u^*}{\text{d}y^*}.
\end{equation}
Hence,
\begin{equation}
u^* = \left\{
\begin{array}{ll}
B~\vert y^* \vert - \frac{G^*}{2} {y^*}^2 + A, & \text{iff}~~ \vert y^* \vert > y^*_p, \\[2pt]
U_p^*, & \text{iff}~~ \vert y^* \vert \leqslant y^*_p,
\end{array} \right.
\end{equation}
where $y^*_p = \frac{B}{G^*}$ and $u(y^*_p) = U^*_p$. The boundary condition for velocity is:
\begin{equation}
u^* \left( \pm \frac{1}{2} \right) = u_s^* = \frac{\beta_s G^*}{2} \left( 1 - \frac{2 B_s}{G^*} \right)
\end{equation}
therefore,
\begin{equation}
u^* = \left\{
\begin{array}{ll}
B~\left( \vert y^* \vert - \displaystyle\frac{1}{2} \right) + \displaystyle\frac{G^*}{2} \left[ \frac{1}{4} - {y^*}^2 + \beta_s \left( 1-\frac{2 B_s}{G^*} \right) \right], & \text{iff}~~ \vert y^* \vert > y^*_p, \\[2pt]
U^*_p, & \text{iff}~~ \vert y^* \vert \leqslant y^*_p.
\end{array} \right.
\end{equation}
Please note that still $G^*$ is unknown and should be calculated from continuity:
\begin{equation}
1 = 2 \int_0^{1/2} u^* ~\text{d}y^* = G^* \left( \frac{1}{12} + \frac{\beta_s}{2} \right) + B \left( \frac{B^2}{3~{G^*}^2} - \frac{1}{4} \right) - \beta_s B_s.
\end{equation}
The individual dissipation terms can be calculated as:
\begin{equation}
a(\boldsymbol{u}^*,\boldsymbol{u}^*) = \int_{-1/2}^{1/2} \left( \frac{\text{d}u^*}{\text{d}y^*} \right)^2 \text{d}y^* = \frac{2}{3G^*} \left( \frac{G^*}{2} - B \right)^3,
\end{equation}

\begin{equation}
B j(\boldsymbol{u}^*) = \int_{-1/2}^{1/2} \vert \frac{\text{d}u^*}{\text{d}y^*} \vert ~\text{d}y^* = B \left( \frac{G^*}{4} - B + \frac{B^2}{G^*} \right),
\end{equation}
and,
\begin{equation}
\int \vert \sigma_{nt}^* ~u_s^* \vert ~\text{d}S = \frac{\beta_s {G^*}^2}{2} \left( 1 - \frac{2 B_s}{G^*} \right).
\end{equation}
However, in the case of fully sliding plug, the slip velocity is,
\begin{equation}
u_s^* = 1 = \frac{\beta_s G^*}{2} \left( 1 - \frac{2 B_s}{G^*} \right).
\end{equation}
Hence,
\begin{equation}
G^* = \frac{2 \left( 1 + \beta_s B_s \right)}{\beta_s} = \frac{2 \left( 1 + \alpha_s \beta_s B \right)}{\beta_s} .
\end{equation}
Therefore, the critical Bingham number beyond which the fully sliding plug occurs is,
\begin{equation}
\bar{B} = \frac{1}{\beta_s (1-\alpha_s)}.
\end{equation}

\section{Flow rate curve in porous media}\label{sec:flowcurve}

\cite{bauer2019experimental} considered 3D flow in model and random porous media using regularized Lattice-Boltzmann approach. The model geometry used was a FCC maximum-packing arrangement of spheres whereas, a random array of overlapping spheres was designed for more realistic porous media. Interestingly, \cite{bauer2019experimental} reported that the model and random porous media display different flow rate dependencies: i.e. flow rate ($\sim B^{-1}$) as a function of the excessive pressure gradient ($\sim Od^{-1} - Od_c^{-1}$) or,
\begin{equation}
B^{-1} \sim \left( Od^{-1} - Od_c^{-1} \right)^{\zeta}.
\end{equation}
They observed that the model porous media predicts a single scaling $\zeta \approx 2$ while the random porous media exhibits two distinctive scalings in the range of moderate and low/high excessive pressure gradients. In other words, in random porous media, both viscous limit and yield (i.e. plastic) limit shares similar characteristics/slopes in predicting the flow rate as a function of the excessive pressure gradient ($\zeta \approx 2$); however, in the viscoplastic regime, the slope is higher ($\zeta \approx 3$). This has been approved by experimental data collected by considering Carbopol flow through sandstone bed.

Using the computational data of the present study under the no-slip condition, we consider how the 2D model porous media and the 2D randomized porous media predicts the flow rate curves and how it differs with \cite{bauer2019experimental} simulations/experiments. Moreover, we investigate if the model and randomized porous media shares the same characteristic in converging to the yield limit at large $B$.

\begin{figure}
\centerline{\includegraphics[width=\linewidth]{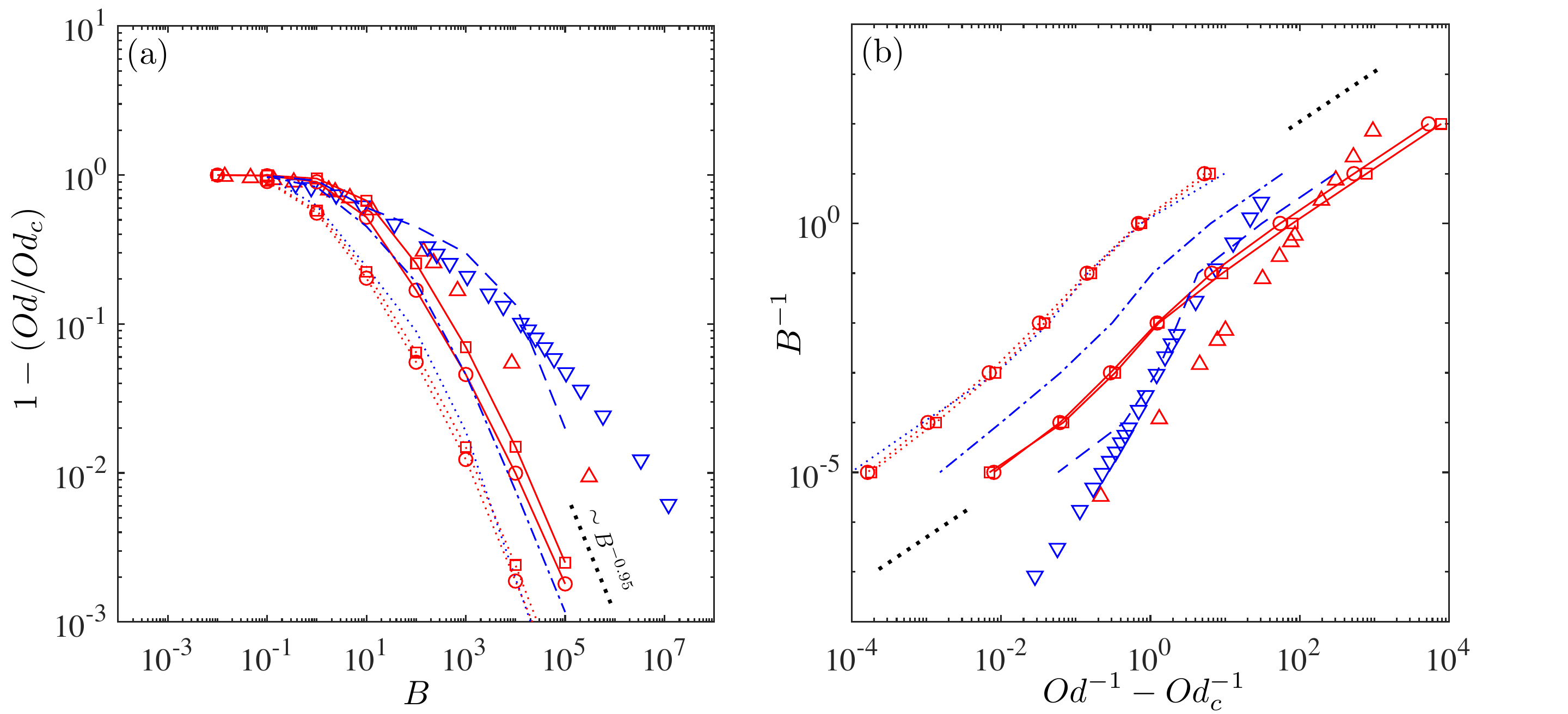}}
\caption{Flow features in model and random-designed porous media. The lines and symbols interpretation is the same as figures \ref{fig:dpdxnoslip} and \ref{fig:dpdxslip}. Please note that in both panels, the lines with circles correspond to the regular model geometry (figure \ref{fig:SchematicPorous}(a)) and those with squares to the staggered model geometry (figure \ref{fig:SchematicPorous}(b)). The dotted black line in panel (a) marks $B^{-0.95}$ scaling and the two dotted black lines in panel (b) 1:1 scaling for reference.}
\label{fig:flowcurve}
\end{figure}

Figure \ref{fig:flowcurve}(a) illustrates the convergence to the yield limit ($1-\frac{Od}{Od_c}$) versus the Bingham number. All the simulations display more or less same scalings:
\begin{equation}
1-\frac{Od}{Od_c} \sim B^{-\nu}
\end{equation}
where $\nu \approx -1$. However, two more conclusions can be drawn from figure \ref{fig:flowcurve}(a):
\begin{itemize}
\item[(i)] At small porosities, we need to go to the higher Bingham numbers to get close enough to the yield limit. This can be attributed to the highly localized and heterogenous/channelized flow at small porosities. Data borrowed from \cite{bauer2019experimental} also show same characteristics, yet a slower convergence rate ($\vert \nu \vert < 0.9$) with the Bingham number compared to our results, which could be the consequence of 3D flows.
\item[(ii)] At high porosities, the model and randomized porous media display closely matched convergence to the yield limit by increasing the Bingham number. For instance, see the dotted blue line (randomized) and the two dotted red lines decorated with symbols (regular and staggered models with circles and squares). However, at the low porosities, the model porous media displays faster convergence compared to the randomized porous media (compare the red continuous lines+symbols ($\phi=0.38$) with the dashed line ($\phi=0.5$)).
\end{itemize}

Another interesting feature to compare, as discussed above, is the flow rate curve; see panel (b) which shows the flow rate versus the excessive pressure gradient. Contradictory to what has been reported by \cite{bauer2019experimental}, this figure shows that the model porous media (at least in 2D), predicts same behaviour: in a sense that at small and large excessive pressure gradients (yield and Newtonian limits, respectively), the flow rate scales linearly with the excessive pressure gradient; however, at the intermediate regime, the flow rate growth as a function of $\frac{1}{Od}-\frac{1}{Od_c}$ is faster. This behaviour is manifested by the randomized porous media as well. Though, as it is clear from panel (b), $\zeta$ is different in our data and the one reported by \cite{bauer2019experimental}. Indeed, both  \cite{bauer2019experimental} and \cite{chevalier2013darcy} demonstrated that $\zeta$ is close to inverse of the power-law index of the Herschel-Bulkley fluid: $\zeta \approx 1/n$. Considering that in the present problem we have used the Bingham model ($n=1$), it makes sense why we found a different exponent ($\zeta \approx 1$) which is consistent with the $\zeta \approx 1/n$ discussion.

\section{Extensions to the non-ideal cases: $k \neq 1$ \& $n \neq 1$}\label{sec:kneq1}

In the main analyses through out the paper, we have assumed that $k=1$. As discussed in the \S 1 \& 2, this assumption is founded on many experimental observations. However, in this appendix, we intend to expand our understanding for the cases in which $k \neq 1$, and also non-ideal viscoplastic rheology. In the resolution of the presented numerical method, `simple' viscoplastic models can be easily adopted such as Herschel-Bulkley and Casson models: it needs updating the step 7 in the presented algorithm, which is well explained by \cite{huilgol2005application}. For instance, for the Herschel-Bulkley model, it takes the form,
\[
\left\{
\begin{array}{ll}
~~~~~~~~~~~~~\hat{\boldsymbol{q}}^{m+1}=0, & \text{iff}~~ \Vert \hat{\boldsymbol{\Sigma}} \Vert \leqslant \hat{\tau}_y, \\[2pt]
\left( \hat{K} \Vert \hat{\boldsymbol{q}}^{m+1} \Vert^{n-1} + \hat{a} \right) \hat{\boldsymbol{q}}^{m+1} = \left( 1- \displaystyle\frac{\hat{\tau}_y}{\Vert\hat{\boldsymbol{\Sigma}}\Vert} \right) \hat{\boldsymbol{\Sigma}}, & \text{iff}~~ \Vert \hat{\boldsymbol{\Sigma}} \Vert > \hat{\tau}_y.
\end{array} \right.
\]
where $\hat{\boldsymbol{\Sigma}} = \hat{\boldsymbol{T}}^m+\hat{a} ~\hat{\dot{\boldsymbol{\gamma}}} \left(\hat{\boldsymbol{u}}^{m+1}\right)$. In the case of more complex models, e.g.~elastoviscoplastic models, one needs more effort; please see \cite{chaparian2019adaptive}.  

\begin{figure}
\centerline{\includegraphics[width=0.7\linewidth]{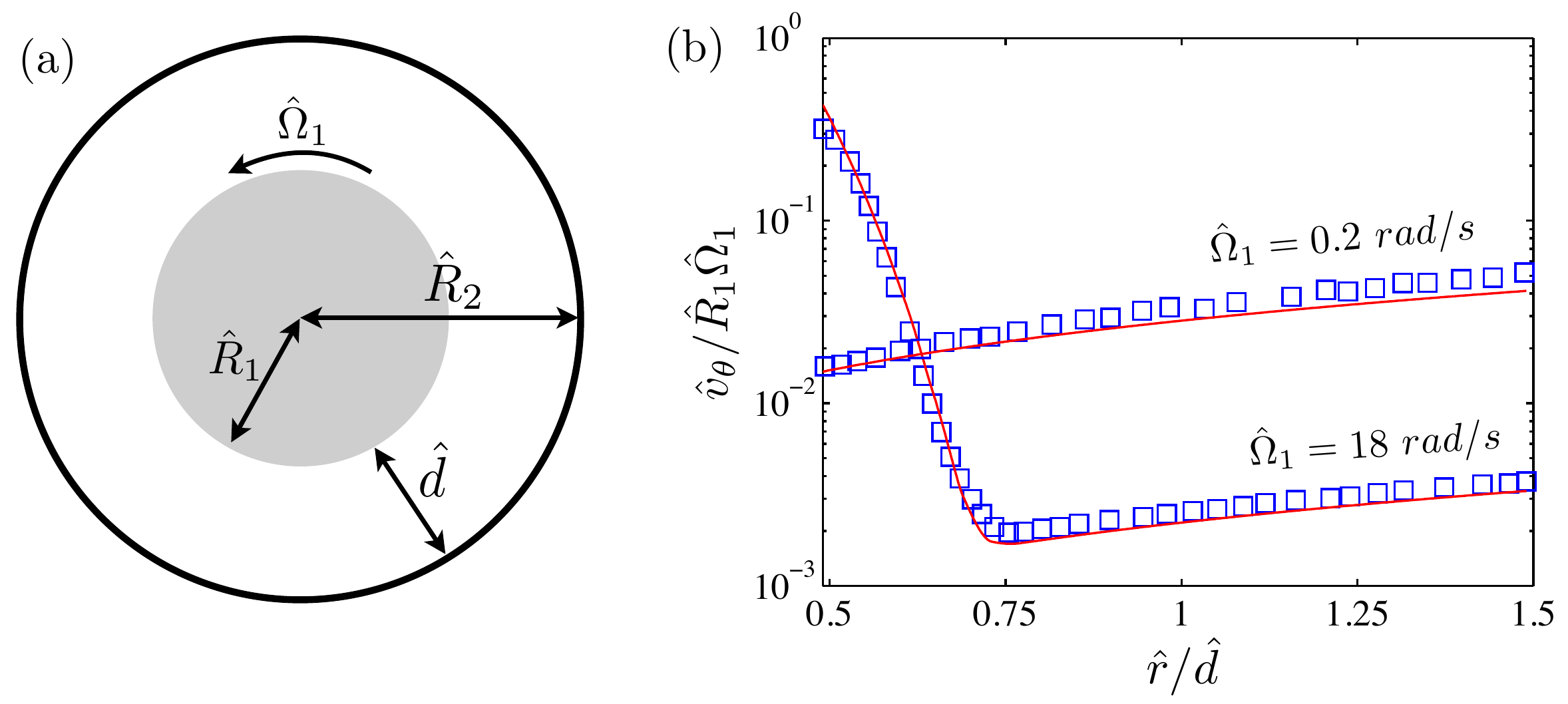}}
\caption{(a) schematic of the 2D Couette rheometry between two cylinders of radii $\hat{R}_1$ (rotating with rotational speed $\hat{\Omega}_1$) and $\hat{R}_2$ (stationary),~(b) red lines are the present computations and the blue symbols are experimental data borrowed form \cite{medina2017tangential}.}
\label{fig:velocity_TC}
\end{figure}

In the following, as an example, we consider the 2D circular Couette flow between two infinitely long cylinders of radii $\hat{R}_1$ and $\hat{R}_2$; see figure \ref{fig:velocity_TC}. We consider this specific geometry to compare our numerical results with the experimental measurements reported by \cite{medina2017tangential}: they reported the velocity profiles in the gap using the Rheo-PIV measurements and also slip laws on the cylinder surfaces for a Carbopol microgel 0.12 wt.~\% with the rheological properties $\hat{\tau}_y = 27 ~Pa,~\hat{K}=5.5~Pa.s^n$ and $n=0.43$. Figure \ref{fig:velocity_TC}(b) compares our numerically obtained velocity profiles (red lines) with \cite{medina2017tangential} experimental measurements for the case $\hat{R}_i = 14 ~mm ~\&~ \hat{R}_o = 42.5 ~mm$ for two different rotational speeds. The slip laws in this case are,
\[
\hat{u}_s = 2.74 \times 10^{-8} ~\vert \hat{\tau}_w \vert^{3.74}
\]
for the deforming regime and,
\[
\hat{u}_s = 5.2 \times 10^{-5} ~\vert \hat{\tau}_w \vert^{1.45}
\]
in the case of sliding plug. As can be seen, there is a close agreement between our simulations and experimental measurements. In the case of $\hat{\Omega}_1 = 18 ~rad/s$, the fluid is yielded close to the inner cylinder (up to $\approx$ 0.75 $\hat{r}/\hat{d}$) and slides over the outer cylinder as a rotating plug, whereas in the case $\hat{\Omega}_1 = 0.2 ~rad/s$ the whole gap is unyielded and undergoes a solid-body rotation.

The only change that we need to implement in the Algorithm \ref{alg2} is the step 8:
\[
\hat{\boldsymbol{\xi}}^{m+1} \gets \displaystyle \left( \hat{\boldsymbol{u}}_{{ns}} \boldsymbol{\cdot} \boldsymbol{n} \right) \boldsymbol{n} + \left( \hat{\boldsymbol{u}}_{{ns}} \boldsymbol{\cdot} \boldsymbol{t} \right) \boldsymbol{t} + \frac{\hat{\beta}_s}{1+ b} ~\hat{\Phi}~\boldsymbol{t}
\]
where $\hat{\Phi} =  - \left( \hat{\boldsymbol{\lambda}}^m \boldsymbol{\cdot} \boldsymbol{t} \right) \vert \hat{\boldsymbol{\lambda}}^m \boldsymbol{\cdot} \boldsymbol{t} \vert^{k-1} + \displaystyle \frac{b}{\hat{\beta}_s}~ \left( \boldsymbol{\delta} \hat{\boldsymbol{u}}^{m+1} \boldsymbol{\cdot} \boldsymbol{t} - \hat{\boldsymbol{u}}_{ns} \boldsymbol{\cdot} \boldsymbol{t} \right)  $.

As discussed in \S \ref{sec:SD}, although hydrodynamics features of the flows are different for different $n$ and $k$, yield limit of the problems are not dependent on these power indices. We will discuss this in the next few lines by considering a sample example in the circular Couette flow configuration and then will generalise this conclusion.

If the inner cylinder is slippery and the outer cylinder is jagged/treated to eliminate slip, then at the yield limit,
\[
B = \frac{\hat{\tau}_y}{\hat{K}} \left( \frac{\hat{d}}{\hat{R}_1 \hat{\Omega}_1} \right)^n \to \infty~~\text{or}~~Y = \frac{\hat{\tau}_y \hat{R}_1^2}{\hat{T}} \to Y_c^-,
\]
the material fully slides over the inner surface and the whole gap is unyielded. Here, $\hat{T}$ denotes the applied torque on the inner cylinder per its unit length. Hence, in the [R] formulation where the scale of the velocity is $\hat{R}_1 \hat{\Omega}_1$ and $\hat{d}$ is the length scale,
\begin{equation}
u_s^* (r=R_1) = 1 = \beta_s \left(\tau_i^* - B_s \right)^k = \beta_s \left(\tau_i^* - \alpha_s B \right)^k,
\end{equation}
and,
\begin{equation}
T^* = 2 \pi \left( \sqrt[k]{\frac{1}{\beta_s}} + \alpha_s B \right).
\end{equation}

Considering the map between the Bingham and the yield number,
\[
Y = B \displaystyle \frac{\hat{\mu} \hat{R}_1^3 \hat{\Omega}_1}{\hat{T} \hat{d}} = \frac{B}{T^*} R_1^2,
\]
we can conclude that the critical yield number is $Y_c = \frac{R_1^2}{2 \pi \alpha_s}$. Hence, the yield limit is independent of $n$ and $k$. This has been demonstrated in the problems studied in the present study that viscous dissipation is at least one order of magnitude less than the plastic dissipation at the yield limit. \cite{frigaard2019background} has shown this before more generally. We have also demonstrated that `viscous' slip dissipation is much less than the `plastic' viscous dissipation at the yield limit for $k=1$. This statement is generally acceptable for $k \neq 1$. Considering that the total slip dissipation term can be split to, 
\begin{equation}
 \int_{\Gamma} \vert \sigma_{nt}^* ~u_s^* \vert ~\text{d}S = \frac{1}{\beta_s^k} \int_{\Gamma}  {u^*_s}^{\frac{1}{k}+1} ~\text{d}S + B_s \int_{\Gamma} \vert u_s^* \vert ~\text{d}S,
\end{equation}
it can be easily understood why the yield limit is independent of $k$: since the `plastic' slip disspation term which is dominant at the yield limit is independent of $k$. Definition (\ref{eq:OdcDef}) is also valid in the case $k \neq 1$ and one can extract the critical ratio of the yield stress to the {\it driving} stress by knowing $\hat{\tau}_s$ or alternatively $\alpha_s$ from the slip law.

\bibliographystyle{jfm}
\bibliography{viscoplastic}

\begin{thebibliography}{55}
\expandafter\ifx\csname natexlab\endcsname\relax\def\natexlab#1{#1}\fi
\def\au#1{#1} \def\ed#1{#1} \def\yr#1{#1}\def\at#1{#1}\def\jt#1{\textit{#1}}
  \def\bt#1{#1}\def\bvol#1{\textbf{#1}} \def\vol#1{#1} \def\pg#1{#1}
  \def\publ#1{#1}\def\arxiv#1{#1}\def\org#1{#1}\def\st#1{\textit{#1}}

\bibitem[Bauer {\em et~al.\/}(2019)Bauer, Talon, Peysson, Ly, Bat{\^o}t,
  Chevalier \& Fleury]{bauer2019experimental}
{\sc \au{Bauer, D.}, \au{Talon, L.}, \au{Peysson, Y.}, \au{Ly, H.~B.},
  \au{Bat{\^o}t, G.}, \au{Chevalier, T.} \& \au{Fleury, M.}} \yr{2019}
  \at{Experimental and numerical determination of {D}arcy's law for yield
  stress fluids in porous media}.  \jt{Phys. Rev. Fluids}  \bvol{4}~(6),
  \pg{063301}.

\bibitem[Beris {\em et~al.\/}(1985)Beris, Tsamopoulos, Armstrong \&
  Brown]{beris1985creeping}
{\sc \au{Beris, A.~N.}, \au{Tsamopoulos, J.~A.}, \au{Armstrong, R.~C.} \&
  \au{Brown, R.~A.}} \yr{1985}  \at{Creeping motion of a sphere through a
  {B}ingham plastic}.  \jt{J. Fluid Mech.}  \bvol{158},  \pg{219--244}.

\bibitem[Chakrabarty(2012)]{chakrabarty2012theory}
{\sc \au{Chakrabarty, J.}} \yr{2012} {\em Theory of plasticity\/}.
  \publ{Butterworth-Heinemann}.

\bibitem[Chaparian \& Frigaard(2017{\natexlab{{\em
  a\/}}})]{chaparian2017cloaking}
{\sc \au{Chaparian, E.} \& \au{Frigaard, I.~A.}} \yr{2017{\natexlab{{\em
  a\/}}}}  \at{Cloaking: Particles in a yield-stress fluid}.  \jt{J.
  Non-Newtonian Fluid Mech.}  \bvol{243},  \pg{47--55}.

\bibitem[Chaparian \& Frigaard(2017{\natexlab{{\em b\/}}})]{chaparian2017yield}
{\sc \au{Chaparian, E.} \& \au{Frigaard, I.~A.}} \yr{2017{\natexlab{{\em
  b\/}}}}  \at{Yield limit analysis of particle motion in a yield-stress
  fluid}.  \jt{J. Fluid Mech.}  \bvol{819},  \pg{311--351}.

\bibitem[Chaparian {\em et~al.\/}(2020)Chaparian, Izbassarov, De~Vita, Brandt
  \& Tammisola]{chaparian2019porous}
{\sc \au{Chaparian, E.}, \au{Izbassarov, D.}, \au{De~Vita, F.}, \au{Brandt, L.}
  \& \au{Tammisola, O.}} \yr{2020}  \at{Yield-stress fluids in porous media: a
  comparison of viscoplastic and elastoviscoplastic flows}.  \jt{Meccanica}
  \bvol{55},  \pg{331--342}.

\bibitem[Chaparian \& Nasouri(2018)]{chaparian2018box}
{\sc \au{Chaparian, E.} \& \au{Nasouri, B.}} \yr{2018}  \at{L-box---a tool for
  measuring the ``yield stress": A theoretical study}.  \jt{Phys. Fluids}
  \bvol{30}~(8),  \pg{083101}.

\bibitem[Chaparian \& Tammisola(2019)]{chaparian2019adaptive}
{\sc \au{Chaparian, E.} \& \au{Tammisola, O.}} \yr{2019}  \at{An adaptive
  finite element method for elastoviscoplastic fluid flows}.  \jt{J.
  Non-Newtonian Fluid Mech.}  \bvol{271},  \pg{104148}.

\bibitem[Chaparian {\em et~al.\/}(2018)Chaparian, Wachs \&
  Frigaard]{chaparian2018inline}
{\sc \au{Chaparian, E.}, \au{Wachs, A.} \& \au{Frigaard, I.~A.}} \yr{2018}
  \at{Inline motion and hydrodynamic interaction of 2{D} particles in a
  viscoplastic fluid}.  \jt{Phys. Fluids}  \bvol{30}~(3),  \pg{033101}.

\bibitem[Chevalier {\em et~al.\/}(2013)Chevalier, Chevalier, Clain, Dupla,
  Canou, Rodts \& Coussot]{chevalier2013darcy}
{\sc \au{Chevalier, T.}, \au{Chevalier, C.}, \au{Clain, X.}, \au{Dupla, J.~C.},
  \au{Canou, J.}, \au{Rodts, S.} \& \au{Coussot, P.}} \yr{2013}  \at{Darcy's
  law for yield stress fluid flowing through a porous medium}.  \jt{J.
  Non-Newtonian Fluid Mech.}  \bvol{195},  \pg{57--66}.

\bibitem[Christel {\em et~al.\/}(2012)Christel, Yahya, Albert \&
  Antoine]{christel2012stick}
{\sc \au{Christel, M.}, \au{Yahya, R.}, \au{Albert, M.} \& \au{Antoine, B.~A.}}
  \yr{2012}  \at{Stick-slip control of the {C}arbopol microgels on polymethyl
  methacrylate transparent smooth walls}.  \jt{Soft Matter}  \bvol{8}~(28),
  \pg{7365--7367}.

\bibitem[Damianou {\em et~al.\/}(2019)Damianou, Panaseti \&
  Georgiou]{damianou2019viscoplastic}
{\sc \au{Damianou, Y.}, \au{Panaseti, P.} \& \au{Georgiou, G.~C.}} \yr{2019}
  \at{Viscoplastic {C}ouette flow in the presence of wall slip with non-zero
  slip yield stress}.  \jt{Mater.}  \bvol{12}~(21),  \pg{3574}.

\bibitem[Daneshi {\em et~al.\/}(2019)Daneshi, Pourzahedi, Martinez \&
  Grecov]{daneshi2019characterising}
{\sc \au{Daneshi, M.}, \au{Pourzahedi, A.}, \au{Martinez, D.~M.} \& \au{Grecov,
  D.}} \yr{2019}  \at{Characterising wall-slip behaviour of {C}arbopol gels in
  a fully-developed {P}oiseuille flow}.  \jt{J. Non-Newtonian Fluid Mech.}
  \bvol{269},  \pg{65--72}.

\bibitem[De~Vita {\em et~al.\/}(2018)De~Vita, Rosti, Izbassarov, Duffo,
  Tammisola, Hormozi \& Brandt]{de2018elastoviscoplastic}
{\sc \au{De~Vita, F.}, \au{Rosti, M.~E.}, \au{Izbassarov, D.}, \au{Duffo, L.},
  \au{Tammisola, O.}, \au{Hormozi, S.} \& \au{Brandt, L.}} \yr{2018}
  \at{Elastoviscoplastic flows in porous media}.  \jt{J. Non-Newtonian Fluid
  Mech.}  \bvol{258},  \pg{10--21}.

\bibitem[Dubash {\em et~al.\/}(2009)Dubash, Balmforth, Slim \&
  Cochard]{dubash2009final}
{\sc \au{Dubash, N.}, \au{Balmforth, N.~J.}, \au{Slim, A.~C.} \& \au{Cochard,
  S.}} \yr{2009}  \at{What is the final shape of a viscoplastic slump?}  \jt{J.
  Non-Newtonian Fluid Mech.}  \bvol{158}~(1-3),  \pg{91--100}.

\bibitem[Fraggedakis {\em et~al.\/}(2016)Fraggedakis, Dimakopoulos \&
  Tsamopoulos]{fraggedakis2016soft}
{\sc \au{Fraggedakis, D.}, \au{Dimakopoulos, Y.} \& \au{Tsamopoulos, J.}}
  \yr{2016}  \at{Yielding the yield-stress analysis: a study focused on the
  effects of elasticity on the settling of a single spherical particle in
  simple yield-stress fluids}.  \jt{Soft Matter}  \bvol{12}~(24),
  \pg{5378--5401}.

\bibitem[Frigaard(2019)]{frigaard2019background}
{\sc \au{Frigaard, I.~A.}} \yr{2019}  \at{Background lectures on ideal
  visco-plastic fluid flows}.  \bt{In {\em Lectures on Visco-Plastic Fluid
  Mechanics\/}},  \pg{pp. 1--40}.  \publ{Springer}.

\bibitem[Hatzikiriakos(2012)]{hatzikiriakos2012wall}
{\sc \au{Hatzikiriakos, S.~G.}} \yr{2012}  \at{Wall slip of molten polymers}.
  \jt{Prog. Polym. Sci.}  \bvol{37}~(4),  \pg{624--643}.

\bibitem[Hecht(2012)]{MR3043640}
{\sc \au{Hecht, F.}} \yr{2012}  \at{New development in freefem++}.  \jt{J.
  Numer. Math.}  \bvol{20}~(3),  \pg{251--265}.

\bibitem[Hewitt \& Balmforth(2018)]{hewitt2018viscoplastic}
{\sc \au{Hewitt, D.~R.} \& \au{Balmforth, N.~J.}} \yr{2018}  \at{Viscoplastic
  slender-body theory}.  \jt{J. Fluid Mech.}  \bvol{856},  \pg{870--897}.

\bibitem[Hill(1950)]{hill1998mathematical}
{\sc \au{Hill, R.}} \yr{1950} {\em The mathematical theory of plasticity\/}.
  \publ{Oxford university press}.

\bibitem[Huilgol(1998)]{huilgol1998variational}
{\sc \au{Huilgol, R.~R.}} \yr{1998}  \at{Variational principle and variational
  inequality for a yield stress fluid in the presence of slip}.  \jt{J.
  Non-Newtonian Fluid Mech.}  \bvol{75}~(2-3),  \pg{231--251}.

\bibitem[Huilgol \& You(2005)]{huilgol2005application}
{\sc \au{Huilgol, Raja~Ramesh} \& \au{You, Z}} \yr{2005}  \at{Application of
  the augmented {L}agrangian method to steady pipe flows of {B}ingham, {C}asson
  and {H}erschel--{B}ulkley fluids}.  \jt{J. Non-Newtonian Fluid Mech.}
  \bvol{128}~(2-3),  \pg{126--143}.

\bibitem[Iglesias {\em et~al.\/}(2020)Iglesias, Mercier, Chaparian \&
  Frigaard]{iglesias2020computing}
{\sc \au{Iglesias, J.~A.}, \au{Mercier, G.}, \au{Chaparian, E.} \&
  \au{Frigaard, I.~A.}} \yr{2020}  \at{Computing the yield limit in
  three-dimensional flows of a yield stress fluid about a settling particle}.
  \jt{arXiv preprint arXiv:2002.05557} .

\bibitem[Izbassarov {\em et~al.\/}(2018)Izbassarov, Rosti, Ardekani, Sarabian,
  Hormozi, Brandt \& Tammisola]{izbassarov2018computational}
{\sc \au{Izbassarov, D.}, \au{Rosti, M.~E.}, \au{Ardekani, M.~N.},
  \au{Sarabian, M.}, \au{Hormozi, S.}, \au{Brandt, L.} \& \au{Tammisola, O.}}
  \yr{2018}  \at{Computational modeling of multiphase viscoelastic and
  elastoviscoplastic flows}.  \jt{Int. J. Numer. Methods Fluids}
  \bvol{88}~(12),  \pg{521--543}.

\bibitem[Jossic \& Magnin(2001)]{jossic2001}
{\sc \au{Jossic, L.} \& \au{Magnin, A.}} \yr{2001}  \at{Drag and stability of
  objects in a yield stress fluid}.  \jt{AIChE J.}  \bvol{47}~(12),
  \pg{2666--2672}.

\bibitem[Liu {\em et~al.\/}(2019)Liu, De~Luca, Rosso \& Talon]{liu2019darcy}
{\sc \au{Liu, C.}, \au{De~Luca, A.}, \au{Rosso, A.} \& \au{Talon, L.}}
  \yr{2019}  \at{Darcy's law for yield stress fluids}.  \jt{Phys. Rev. Lett.}
  \bvol{122}~(24),  \pg{245502}.

\bibitem[Martin \& Randolph(2006)]{martin2006upper}
{\sc \au{Martin, C.~M.} \& \au{Randolph, M.~F.}} \yr{2006}  \at{Upper-bound
  analysis of lateral pile capacity in cohesive soil}.  \jt{G{\'e}otechnique}
  \bvol{56}~(2),  \pg{141--145}.

\bibitem[Medina-Ba{\~n}uelos {\em et~al.\/}(2019)Medina-Ba{\~n}uelos,
  Mar{\'\i}n-Santib{\'a}{\~n}ez, P{\'e}rez-Gonz{\'a}lez \&
  Kalyon]{medina2019rheo}
{\sc \au{Medina-Ba{\~n}uelos, E.~F.}, \au{Mar{\'\i}n-Santib{\'a}{\~n}ez,
  B.~M.}, \au{P{\'e}rez-Gonz{\'a}lez, J.} \& \au{Kalyon, D.~M.}} \yr{2019}
  \at{Rheo-{PIV} analysis of the vane in cup flow of a viscoplastic microgel}.
  \jt{J. Rheol.}  \bvol{63}~(6),  \pg{905--915}.

\bibitem[Medina-Ba{\~n}uelos {\em et~al.\/}(2017)Medina-Ba{\~n}uelos,
  Mar{\'\i}n-Santib{\'a}{\~n}ez, P{\'e}rez-Gonz{\'a}lez, Malik \&
  Kalyon]{medina2017tangential}
{\sc \au{Medina-Ba{\~n}uelos, E.~F.}, \au{Mar{\'\i}n-Santib{\'a}{\~n}ez,
  B.~M.}, \au{P{\'e}rez-Gonz{\'a}lez, J.}, \au{Malik, M.} \& \au{Kalyon,
  D.~M.}} \yr{2017}  \at{Tangential annular ({C}ouette) flow of a viscoplastic
  microgel with wall slip}.  \jt{J. Rheol.}  \bvol{61}~(5),  \pg{1007--1022}.

\bibitem[Meeker {\em et~al.\/}(2004{\natexlab{{\em a\/}}})Meeker, Bonnecaze \&
  Cloitre]{meeker2004JOR}
{\sc \au{Meeker, S.~P.}, \au{Bonnecaze, R.~T.} \& \au{Cloitre, M.}}
  \yr{2004{\natexlab{{\em a\/}}}}  \at{Slip and flow in pastes of soft
  particles: {D}irect observation and rheology}.  \jt{J. Rheol.}
  \bvol{48}~(6),  \pg{1295--1320}.

\bibitem[Meeker {\em et~al.\/}(2004{\natexlab{{\em b\/}}})Meeker, Bonnecaze \&
  Cloitre]{meeker2004PRL}
{\sc \au{Meeker, S.~P.}, \au{Bonnecaze, R.~T.} \& \au{Cloitre, M.}}
  \yr{2004{\natexlab{{\em b\/}}}}  \at{Slip and flow in soft particle pastes}.
  \jt{Phys. Rev. Lett.}  \bvol{92}~(19),  \pg{198302}.

\bibitem[Mosolov \& Miasnikov(1965)]{mosolov1965variational}
{\sc \au{Mosolov, P.~P.} \& \au{Miasnikov, V.~P.}} \yr{1965}  \at{Variational
  methods in the theory of the fluidity of a viscous-plastic medium}.  \jt{J.
  Appl. Math. Mech.}  \bvol{29}~(3),  \pg{545--577}.

\bibitem[Muravleva(2018)]{muravleva2018squeeze}
{\sc \au{Muravleva, L.}} \yr{2018}  \at{Squeeze flow of {B}ingham plastic with
  stick-slip at the wall}.  \jt{Phys. Fluids}  \bvol{30}~(3),  \pg{030709}.

\bibitem[Murff {\em et~al.\/}(1989)Murff, Wagner \& Randolph]{murff1989pipe}
{\sc \au{Murff, J.~D.}, \au{Wagner, D.~A.} \& \au{Randolph, M.~F.}} \yr{1989}
  \at{Pipe penetration in cohesive soil}.  \jt{G{\'e}otechnique}
  \bvol{39}~(2),  \pg{213--229}.

\bibitem[Nirmalkar {\em et~al.\/}(2012)Nirmalkar, Chhabra \&
  Poole]{nirmalkar2012creeping}
{\sc \au{Nirmalkar, N.}, \au{Chhabra, R.~P.} \& \au{Poole, R.~J.}} \yr{2012}
  \at{On creeping flow of a {B}ingham plastic fluid past a square cylinder}.
  \jt{J. Non-Newtonian Fluid Mech.}  \bvol{171},  \pg{17--30}.

\bibitem[Panaseti \& Georgiou(2017)]{panaseti2017viscoplastic}
{\sc \au{Panaseti, P.} \& \au{Georgiou, G.~C.}} \yr{2017}  \at{Viscoplastic
  flow development in a channel with slip along one wall}.  \jt{J.
  Non-Newtonian Fluid Mech.}  \bvol{248},  \pg{8--22}.

\bibitem[Philippou {\em et~al.\/}(2016)Philippou, Kountouriotis \&
  Georgiou]{philippou2016viscoplastic}
{\sc \au{Philippou, M.}, \au{Kountouriotis, Z.} \& \au{Georgiou, G.~C.}}
  \yr{2016}  \at{Viscoplastic flow development in tubes and channels with wall
  slip}.  \jt{J. Non-Newtonian Fluid Mech.}  \bvol{234},  \pg{69--81}.

\bibitem[Piau(2007)]{piau2007carbopol}
{\sc \au{Piau, J.~M.}} \yr{2007}  \at{Carbopol gels: Elastoviscoplastic and
  slippery glasses made of individual swollen sponges: Meso-and macroscopic
  properties, constitutive equations and scaling laws}.  \jt{J. Non-Newtonian
  Fluid Mech.}  \bvol{144}~(1),  \pg{1--29}.

\bibitem[Poumaere {\em et~al.\/}(2014)Poumaere, Moyers-Gonz{\'a}lez, Castelain
  \& Burghelea]{poumaere2014unsteady}
{\sc \au{Poumaere, A.}, \au{Moyers-Gonz{\'a}lez, M.}, \au{Castelain, C.} \&
  \au{Burghelea, T.}} \yr{2014}  \at{Unsteady laminar flows of a {C}arbopol gel
  in the presence of wall slip}.  \jt{J. Non-Newtonian Fluid Mech.}
  \bvol{205},  \pg{28--40}.

\bibitem[Putz \& Frigaard(2010)]{putz2010creeping}
{\sc \au{Putz, A.} \& \au{Frigaard, I.~A.}} \yr{2010}  \at{Creeping flow around
  particles in a {B}ingham fluid}.  \jt{J. Non-Newtonian Fluid Mech.}
  \bvol{165}~(5),  \pg{263--280}.

\bibitem[Putz {\em et~al.\/}(2008)Putz, Burghelea, Frigaard \&
  Martinez]{putz2008settling}
{\sc \au{Putz, A. M.~V.}, \au{Burghelea, T.~I.}, \au{Frigaard, I.~A.} \&
  \au{Martinez, D.~M.}} \yr{2008}  \at{Settling of an isolated spherical
  particle in a yield stress shear thinning fluid}.  \jt{Phys. Fluids}
  \bvol{20}~(3),  \pg{33102--33300}.

\bibitem[Randolph \& Houlsby(1984)]{randolph1984limiting}
{\sc \au{Randolph, M.~F.} \& \au{Houlsby, G.~T.}} \yr{1984}  \at{The limiting
  pressure on a circular pile loaded laterally in cohesive soil}.
  \jt{G{\'e}otechnique}  \bvol{34}~(4),  \pg{613--623}.

\bibitem[Roquet \& Saramito(2003)]{roquet2003adaptive}
{\sc \au{Roquet, N.} \& \au{Saramito, P.}} \yr{2003}  \at{An adaptive finite
  element method for {B}ingham fluid flows around a cylinder}.  \jt{Comput.
  Meth. Appl. Mech. Eng.}  \bvol{192}~(31),  \pg{3317--3341}.

\bibitem[Roquet \& Saramito(2008)]{roquet2008adaptive}
{\sc \au{Roquet, N.} \& \au{Saramito, P.}} \yr{2008}  \at{An adaptive finite
  element method for viscoplastic flows in a square pipe with stick--slip at
  the wall}.  \jt{J. Non-Newtonian Fluid Mech.}  \bvol{155}~(3),
  \pg{101--115}.

\bibitem[Roustaei {\em et~al.\/}(2016)Roustaei, Chevalier, Talon \&
  Frigaard]{roustaei2016non}
{\sc \au{Roustaei, A.}, \au{Chevalier, T.}, \au{Talon, L.} \& \au{Frigaard,
  I.~A.}} \yr{2016}  \at{Non-{D}arcy effects in fracture flows of a yield
  stress fluid}.  \jt{J. Fluid Mech.}  \bvol{805},  \pg{222--261}.

\bibitem[Roustaei \& Frigaard(2013)]{roustaei2013occurrence}
{\sc \au{Roustaei, A.} \& \au{Frigaard, I.~A.}} \yr{2013}  \at{The occurrence
  of fouling layers in the flow of a yield stress fluid along a wavy-walled
  channel}.  \jt{J. Non-Newtonian Fluid Mech.}  \bvol{198},  \pg{109--124}.

\bibitem[Roustaei \& Frigaard(2015)]{roustaei2015residualb}
{\sc \au{Roustaei, A.} \& \au{Frigaard, I.~A.}} \yr{2015}  \at{Residual
  drilling mud during conditioning of uneven boreholes in primary cementing.
  part 2: Steady laminar inertial flows}.  \jt{J. Non-Newtonian Fluid Mech.}
  \bvol{226},  \pg{1--15}.

\bibitem[Roustaei {\em et~al.\/}(2015)Roustaei, Gosselin \&
  Frigaard]{roustaei2015residuala}
{\sc \au{Roustaei, A.}, \au{Gosselin, A.} \& \au{Frigaard, I.~A.}} \yr{2015}
  \at{Residual drilling mud during conditioning of uneven boreholes in primary
  cementing. part 1: Rheology and geometry effects in non-inertial flows}.
  \jt{J. Non-Newtonian Fluid Mech.}  \bvol{220},  \pg{87--98}.

\bibitem[Seth {\em et~al.\/}(2012)Seth, Locatelli-Champagne, Monti, Bonnecaze
  \& Cloitre]{seth2012soft}
{\sc \au{Seth, J.~R.}, \au{Locatelli-Champagne, C.}, \au{Monti, F.},
  \au{Bonnecaze, R.~T.} \& \au{Cloitre, M.}} \yr{2012}  \at{How do soft
  particle glasses yield and flow near solid surfaces?}  \jt{Soft Matter}
  \bvol{8}~(1),  \pg{140--148}.

\bibitem[Talon \& Bauer(2013)]{talon2013determination}
{\sc \au{Talon, L.} \& \au{Bauer, D.}} \yr{2013}  \at{On the determination of a
  generalized {D}arcy equation for yield-stress fluid in porous media using a
  {L}attice-{B}oltzmann {TRT} scheme}.  \jt{Eur. Phys. J. E}  \bvol{36}~(12),
  \pg{139}.

\bibitem[Tokpavi {\em et~al.\/}(2009)Tokpavi, Jay, Magnin \&
  Jossic]{tokpavi2009experimental}
{\sc \au{Tokpavi, D.~L.}, \au{Jay, P.}, \au{Magnin, A.} \& \au{Jossic, L.}}
  \yr{2009}  \at{Experimental study of the very slow flow of a yield stress
  fluid around a circular cylinder}.  \jt{J. Non-Newtonian Fluid Mech.}
  \bvol{164}~(1),  \pg{35--44}.

\bibitem[Tokpavi {\em et~al.\/}(2008)Tokpavi, Magnin \& Jay]{tokpavi2008very}
{\sc \au{Tokpavi, D.~L.}, \au{Magnin, A.} \& \au{Jay, P.}} \yr{2008}  \at{Very
  slow flow of {B}ingham viscoplastic fluid around a circular cylinder}.
  \jt{J. Non-Newtonian Fluid Mech.}  \bvol{154}~(1),  \pg{65--76}.

\bibitem[Vand(1948)]{vand1948viscosity}
{\sc \au{Vand, V.}} \yr{1948}  \at{Viscosity of solutions and suspensions. {I}.
  {T}heory}.  \jt{J. Phys. Chem.}  \bvol{52}~(2),  \pg{277--299}.

\bibitem[Zhang {\em et~al.\/}(2018)Zhang, Lorenceau, Bourouina, Basset,
  Oerther, Ferrari, Rouyer, Goyon \& Coussot]{zhang2018wall}
{\sc \au{Zhang, X.}, \au{Lorenceau, E.}, \au{Bourouina, T.}, \au{Basset, P.},
  \au{Oerther, T.}, \au{Ferrari, M.}, \au{Rouyer, F.}, \au{Goyon, J.} \&
  \au{Coussot, P.}} \yr{2018}  \at{Wall slip mechanisms in direct and inverse
  emulsions}.  \jt{J. Rheol.}  \bvol{62}~(6),  \pg{1495--1513}.

\end{thebibliography}

\end{document}